\documentclass[fleqn,usenatbib]{mnras}

\usepackage{mathptmx}
\usepackage[T1]{fontenc}

\DeclareRobustCommand{\VAN}[3]{#2}
\let\VANthebibliography\thebibliography
\def\thebibliography{\DeclareRobustCommand{\VAN}[3]{##3}\VANthebibliography}

\usepackage{graphicx}	% Including Fig.~files
\usepackage{amsmath, tabu}	% Advanced maths commands
\usepackage{amssymb}	% Extra maths symbols
\usepackage{adjustbox}
\usepackage{makecell, multirow}
\usepackage{hyperref}

\newcommand{\Msun}{\ensuremath{\,{\rm M}_\odot}}                  % Solar mass symbol
\newcommand{\Rsun}{\ensuremath{\,{\rm R}_\odot}}                  % Solar radius symbol
\newcommand{\psun}{\ensuremath{\,\rho_\odot}}                     % Solar density symbol
\newcommand{\Teff}{\ensuremath{T_{\rm eff}}}
\newcommand{\reff}[1]{#1}

\title[Low mass eclipsing binary stars]{Revising the properties of low mass eclipsing binary stars using TESS light curves}

\author[Z.\ Jennings et al.]{
Z.\ Jennings$^{1}$\thanks{E-mail: z.jennings@keele.ac.uk},
J.\ Southworth$^{1}$,
P.\ F.\ L.\ Maxted$^{1}$,
L.\ Mancini$^{2,3,4,5}$
\\
$^{1}$\,Astrophysics Group, Keele University, Staffordshire, ST5 5BG, UK \\
$^{2}$\,Department of Physics, University of Rome ``Tor Vergata'', Via della Ricerca Scientifica 1, 00133 -– Rome, Italy \\
$^3$\,Max Planck Institute for Astronomy, K\"onigstuhl 17, 69117 –- Heidelberg, Germany \\
$^4$\,INAF –- Osservatorio Astrofisico di Torino, via Osservatorio 20, 10025 –- Pino Torinese, Italy \\
$^5$\,International Institute for Advanced Scientific Studies (IIASS), Via G.\ Pellegrino 19, 84019 –- Vietri sul Mare (SA), Italy
}

\date{Accepted XXX. Received YYY; in original form ZZZ}

\pubyear{2022}

\begin{document}
\label{firstpage}
\pagerange{\pageref{firstpage}--\pageref{lastpage}}
\maketitle

\begin{abstract}
\reff{Precise measurements of stellar parameters are required in order to develop our theoretical understanding of stellar structure. These measurements enable errors and uncertainties to be quantified in theoretical models and constrain the physical interpretation of observed phenomena, such as the inflated radii of low-mass stars}. 

We use newly-available TESS light curves combined with published radial velocity measurements to improve the characterization of 12 low mass eclipsing binaries composed of an M~dwarf accompanied by a brighter F/G star. We present and analyse ground-based simultaneous four-colour photometry for two targets. Our results include the first measurements of the fundamental properties of two of the systems. Light curve and radial velocity information were converted into the physical parameters of each component of the systems using an isochrone fitting method. We also derive the effective temperatures of the M~dwarfs, almost tripling the number of such measurements.

The results are discussed in the context of radius inflation. We find that exquisite precision in the age estimation of young objects is required to determine their inflation status. However, all but three of the objects are securely located among the main sequence, demonstrating radius inflation and the necessity to develop our understanding of the complex physical processes governing the evolution of low-mass stars. \reff{We investigated the hypothesis that luminosity is unaffected by the inflation problem but the findings were not conclusive.}

%Our results based on a sample of M-dwarfs in single-lined binary systems do not support the hypothesis that luminosity is unaffected by the inflation problem. 
\end{abstract}


%%%%%%%%%%%%%%%%%%%%%%%%%%%%%%%%%%%%%%%%%%%%%%%%%%

%%%%%%%%%%%%%%%%% BODY OF PAPER %%%%%%%%%%%%%%%%%%

\section{Introduction}\label{Introduction.S}

Stellar theory is well understood for stars with masses of 1--5\Msun\ \citep{Claret_2021} compared to low-mass stars in the regime of 0.08--0.3\Msun. Such stars are challenging because of the complex and varied physics which occurs in their interiors, in particlar magnetic phenomena \citep{Mullan_2001}. The lowest-mass stars are near the hydrogen burning limit and are partially degenerate \citep{ChabrierBaraffe00ara}.

%For example, they are near the hydrogen burning limit and are cool enough that their interior temperatures are on the order of the electron Fermi temperature, causing parts of their interiors to behave as a partially degenerate gas \citep{Chabrier_1997,Beatty_2007,Fernandez_2009}. The electron number density is such that the mean inter-ionic distance is on the order of the Fermi screening length, meaning the electron gas is polarized by the external field. Attempts to describe the interior of low mass stars must therefore take into account the partially degenerate polarized plasma while further complexity may be introduced by magnetism \citep{Mullan_2001,Beatty_2007}, where the field is not only stronger, but more pervasive \citep{Hawley_2000}. Relatively lower temperatures would also allow for the recombination of molecular hydrogen and other other molecules such as Ti0, meaning that the grey atmosphere approximation becomes least valid as the opacity spectrum becomes most complex. 

Therefore, the interior structure of fully convective low-mass stars is complex and different to that of partially radiative higher-mass stars. A lack of observational constraints on their properties \citep{Swayne_2021} makes it difficult to address inaccuracies in our understanding of their structure. The radii of low-mass stars are observed to be inflated by $\approx 10\%$ \citep{Zhou_2014} compared to theoretical predictions \citep{Hoxie73aa,Lacy77apjs,Lopez07apj,Torres13an}. \reff{The discrepancy has also been observed to persist in stars with masses up to 1\Msun\ \citep[e.g.][]{Torres_2021,Southworth_2022} and the problem may even be greater for early-M dwarfs (see Section\,\ref{Discussion.S})}. 

Most of the objects whose measured radii are precise enough to usefully constrain theoretical models exist in eclipsing binary systems (EBs). It has been suggested that tidal interactions in EBs leads to faster rotation and increased magnetic activity, which decreases the efficiency of convective energy transport, causing the radius to expand \citep{Mullan_2001, Lopez-Morales_Ribas_2005, Lopez-Morales_2007}. Surface activity detectable in some of their light curves \citep[e.g.][]{Morales_2007,Torres_2006} supports this hypothesis and, indeed, artificially low values for the mixing length parameter in the outer convective zone reduce the observed discrepancy considerably \citep{Torres_2006,Chabrier_2007}. However, radius inflation has also been observed for isolated low mass stars, which rotate slowly due to magnetic braking, so explanations should not be restricted to binary systems \citep{Berger+06apj,MorrellNaylor19mn}. Other possible causes such as metallicity and uncertainties in the input physics have been discussed \citep[e.g.][]{Swayne_2021,Torres_2010}. It should be noted that ``radius inflation'' has not been observed in some studies \citep{Zhou_2014, Bentley_2009} to within the measurement errors. 

It is clear that more observational constraints are required in order to accurately address radius inflation and the associated theoretical uncertainties in descriptions of low-mass stellar interiors. It is possible to obtain these measurements to the required precision for EBs that are spectroscopically double-lined (SB2), by modelling their light and radial velocity (RV) curves \citep{Andersen91aarv,Torres_2010}. However, known examples of such systems are relatively rare\footnote{See the Detached Eclipsing Binary Catalogue \citep{Me15aspc} at \texttt{https://www.astro.keele.ac.uk/jkt/debcat/}.} due to the low binary fraction in low-mass stars \citep{DucheneKraus13araa} and their intrinsic faintness.

The advent of wide-field searches for planetary transits has led to the discovery of many eclipsing binaries with low-mass companions (EBLMs) \reff{\citep{Beatty_2007,Fernandez_2009,Triaud_2017, Collins_2018, Zhoe_2015}}, where an M~dwarf transits a much larger and brighter F or G dwarf. The faintness of the M~dwarf secondary stars versus the F/G primary components means they are usually not detected in spectra of the system, making EBLMs single-lined spectroscopic binaries (SB1s). Such systems offer a way to provide precise measurements of the masses and radii of M~dwarfs but the only direct measurements that can be made for these systems are the density of the primary component and the surface gravity of the secondary. An additional constraint is required to set the scale of the system. One approach is to establish the properties of the primary components via isochrone fitting, leading to direct but model-dependent measurements of the properties of the M~dwarfs. Another is to assume that the system is rotationally synchronized. A third is to use empirical relations for solar-type stars to specify the properties of the primary components without using stellar theory \citep{Enoch+10aa,Me10mn,Hartman+15aj}. \reff{Alternatively, given that the parallax, bolometric magnitude and temperatures of the object is measured, the total luminosity can provide the needed constraint.} %This has been taken advantage of by \citet{Fernandez_2009}; \citet{Bentley_2009}; \citet{Koo_2012}; \citet{Latham_2009}; \citet{Zhou_2014}; \citet{Wang_2014}; \citet{Eigmuller_2016}; \citet{Beatty_2007}; \citet{Pasternacki_2011}; \citet{Crouzet_2013}. 

Thirteen EBLMs considered in this study have been observed using ground-based photometry. Such data are sufficient to detect and measure the properties of EBLMs, but the observational scatter often limits the precision of the mass and radius measurements. \reff{The situation can be greatly improved by using space-based photometry due to its competitive precision coupled with improved time sampling over longer periods of continuous monitoring}. The use of photometry from space satellites has revolutionised the study of EBs \citep{Me21univ}. The Transiting Exoplanet Survey Satellite (TESS, \citealt{Ricker+15jatis}) is the most useful mission because it has observed the great majority of the sky in both hemispheres. In this work, we use light curves from TESS and published RVs to obtain new measurements of the physical properties of a set of known EBLMs. The aim is to provide improved constraints on theoretical models of low-mass stars and thus more accurately address the uncertainties surrounding the interior structure of these objects. We also present new ground-based high-precision light curves of two objects obtained in four passbands simultaneously.

Basic information regarding the systems studied in this work are given in Table\,\ref{basic_params.tab}. These objects were chosen with the aim to include all EBLMs with published RVs and previously unstudied TESS light curves. However, we explicitly excluded objects in the EBLM series of papers \citep[see][]{Triaud+17aa} as these are currently being analysed by others.

In Section~\ref{Observations.S} a description of the observations and data used in this work is presented. Section~\ref{Analysis.S} outlines the analysis techniques, and the results for each system are presented in Section~\ref{Results.S}. A discussion and concluding remarks are presented in Section~\ref{Discussion.S}.  

\begin{table*}\caption{Basic parameters for the EBLMs included in this work. Values for \Teff\ were taken from the literature when such analysis was reliable. In other cases, indicated with an asterisk next to the \Teff, the values were determined from a SED fit as described in Section \ref{sec:physical_properties}.}
    \begin{adjustbox}{width=1\textwidth}
    \centering
    \begin{tabular}{lrlcccc}
    \hline 
    \multicolumn{3}{c}{\multirow{1}{*}{{Identifiers}}} & \multirow{2}{*}{Reference(s)} &  \multirow{2}{*}{$V$ mag} & \multirow{2}{*}{TESS sectors} & Temperature \\ 
    %\cline{1-3}
     \multicolumn{1}{c}{This work} & TIC~~~~~~ & Published & & & & (K) \\ \hline
    %\hline 
    {TYC 2755-36-1} & 305982045 & HAT-TR-205-013 &  \citet{Beatty_2007, Latham_2009} & 10.72 & -- & \makebox[1.3cm][s]{$6617\pm200^*$}  \\
    {HAT-TR-205-003} & 115686059 & HAT-TR-205-003 & \citet{Latham_2009} & 12.48 & 16 & \makebox[1.3cm][s]{$6363\pm150^*$}  \\
    {T-Aur0-13378} & 122104276 & T-Aur0-13378 &  \citet{Fernandez_2009} & 13.35 & 19 & \makebox[1.3cm][s]{$6675\pm125^*$}  \\
    {TYC 3576-2035-1} & 295803225 & T-Cyg1-01385  & \citet{Fernandez_2009} &10.70 & 14, 15 & \makebox[1.3cm][s]{$5887\pm125^*$}  \\
    {TYC 3473-673-1} & 148781497 & T-Boo0-00080  & \citet{Fernandez_2009}& 10.30 & 16, 23 & \makebox[1.3cm][s]{$6254\pm125^*$} \\
    {TYC 3545-371-1} & 48450535 & T-Lyr1-01662 &  \citet{Fernandez_2009} & 11.30 & 14, 15 & \makebox[1.3cm][s]{$6956\pm125^*$}  \\
    {TYC 3121-1659-1} & 394178587 & T-Lyr0-08070 &  \citet{Fernandez_2009} & 12.30 & 14 & \makebox[1.3cm][s]{$6633\pm125^*$}  \\
    {TYC 7096-222-1} & 53059882 & TYC 7096-222-1 &  \citet{Bentley_2009} & 10.28 & 6, 7, 33, 34 & $7600\pm300$ \\
    {TYC 2855-585-1} & 192587088 & TYC 2855-585-1   & \citet{Koo_2012} & 11.31 & 18 & $6500 \pm 250$ \\
    {TYC 9535-351-1} & 451596415 & TYC 9535-351-1 & \citet{Wang_2014,Crouzet_2013} &10.12&13,27,39&\makebox[1.3cm][s]{$6663\pm125^*$} \\
    {TYC 6493-290-1} & 32606889 & HATS 551-019  & \citet{Zhou_2014} & 12.06 & 5, 6, 32, 33 & $6380\pm170$ \\
    {GSC 06493-00315} & 32677675 & HATS 551-021 & \citet{Zhou_2014} & 13.11 & 5, 6, 32, 33 & $6670\pm220$ \\
    {GSC 05946-00892} & 59751429 & HATS 553-001 & \citet{Zhou_2014} & 13.19 & 33 & $6230\pm250$  \\
    {GSC 06465-00602} & 594723 & HATS 550-016 & \citet{Zhou_2014} & 13.61 & 32 & $6420\pm90$  \\
    {TYC 3700-1739-1} & 645967562 & TYC 3700-1739-1 & \citet{Eigmuller_2016} & 11.77 & 18 & $7350\pm250$ \\
    \hline
    \end{tabular}
    \end{adjustbox}
    \label{basic_params.tab}
\end{table*}

\section{Observations}\label{Observations.S}

\subsection{TESS observations}

TESS has observed over 200\,000 selected main sequence dwarfs in 2-minute cadence (SC) mode and many more in the 30-minute cadence (LC) mode in which the data are saved as a full-frame-image (FFI). We account for the sparser sampling of the latter using numerical integration (see Section\,\ref{Analysis.S}).

%, which uses 2 minute observations, and this number is multiplied by several factors when including objects observed using their 30 minute long cadence (LC) mode. Long cadence observations use the entire, four camera field of view of the TESS instrument and the corresponding data is stored in a full frame image (FFI). Light curves have been obtained for the targets in Table\,\ref{basic_params.tab} by TESS in either SC or LC except for TYC 2755-36-1, where ground based data of sufficient quality is used for its characterization.
% For the target TYC 3121-1659-1, there 
% were significantly less data points in the TESS observations and so ground based
% data of similar quality is added analysis of this star.     

The Web TESS viewing tool\footnote{\texttt{https://heasarc.gsfc.nasa.gov/cgi-bin/tess/webtess/wtv.py}} was used to determine which and how many sectors the targets had been observed in. For SC data, both simple aperture photometric (SAP) and pre-search data conditioned SAP (PDCSAP) data \citep{Jenkins+16spie} were inspected before choosing one of them as the most suitable. This is necessary since the PDC reduction pipeline usually yields more precise data but can introduce unwanted artefacts in targets dissimilar to those that the routine is tailored to. 

For targets observed in LC mode, the data were extracted from the FFIs using custom routines. Aperture photometry was performed using apertures whose size and position were adjusted manually for each target to optimally extract the flux of each target while minimsing background flux from neighbours. The surrounding field was investigated by first querying the \textit{Vizier}\footnote{\texttt{https://vizier.cds.unistra.fr/}} database for all objects within a 3.5 arcminute radius from the target with an apparent \textit{Gaia} $G$ magnitude of less than 16 in the \textit{Gaia} DR2 catalogue \citep{Gaia18aa}. A threshold magnitude of $G=16$ was chosen since objects fainter than this are not expected to be bright enough to have any effect on the observations \citep{Me+20aa}. The positions of these objects were then marked on the 200th frame of the target, chosen due to early frames in the time series being contaminated by scattered light from Earth at perigee. Verifying the locations of neighbouring objects also aided the choice of the size and location of the aperture. The positions of any marked contaminants in \textit{Gaia} DR2 were further confirmed using the Two Micron All Sky Survey (2MASS; \citealt{2MASS}) as a reference when available. In a few cases where a 2MASS image was unavailable, we used images from the ESO Digitized Sky Survey (DSS) instead. The level of contaminant flux captured within the aperture of LC targets was investigated by plotting the position of the centroid of the target in the x and y planes of the target pixel file against time. Shifts in the position of the centroids occur during eclipses when contamination is serious.

All LC and SC data were normalized by dividing the flux and error by the median flux value. Quasi-periodic variations present in most light curves were attributed to starspots in the primary component and divided out. Before dividing out the magnetic activity, it was necessary to mask the eclipses. Astropy's implementation of the box least squares algorithm \citep{Astropy1,Astropy2}, which models a transit as a periodic upside-down top hat with four simple parameters, was used to develop a mask for the primary eclipses. The parameters of the model are the transit midpoint, duration, period and depth, where the first three of these were used in the masking process. The secondary eclipses were masked manually. 

The variations were then modelled via two methods and divided out. The most effective method based on a visual inspection of the resulting light curve was chosen. The first method used Astropy's implementation of Lomb-Scargle periodograms in order to model the observed variations with a combination of sinusoids. The second method applied a Savitsky-Golay filter, as implemented by the python package \textsc{lightkurve} \citep{lightkurve}, to the masked time series, fitting successive subsets of adjacent datapoints with a low-degree polynomial by linear least squares. Trial values for the degree of polynomial were used and the best result carried forward for comparison with the first method. 

\begin{figure*}
    \centering
    \includegraphics[width = \textwidth]{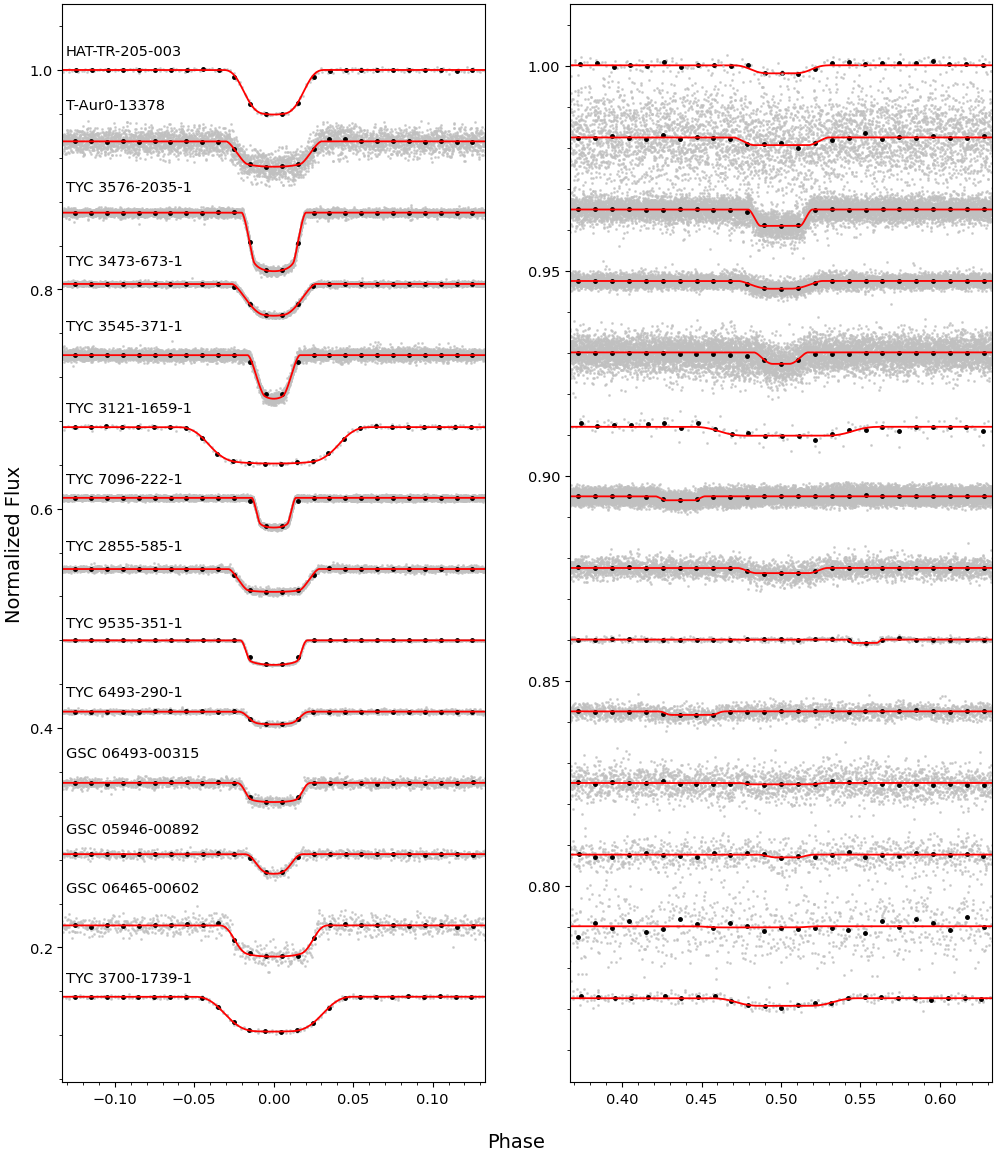}
    \caption{The TESS light curves for our targets around primary eclipse (left) and secondary eclipse (right) compared to the fitted model (lines). \reff{Binned data (black) is plotted over the raw (grey) data}. The system TYC 2755-36-1 is not included because no TESS data are available for it.}
    \label{Tess.fig}
\end{figure*}

Some objects were observed by TESS in more than one sector. In these cases, the data from each sector were concatenated into a single data file after the above process was carried out individually for each sector. The phase-folded TESS data are shown in Fig.\,\ref{Tess.fig}.

\begin{figure*}
    \centering
    \includegraphics[width = \textwidth]{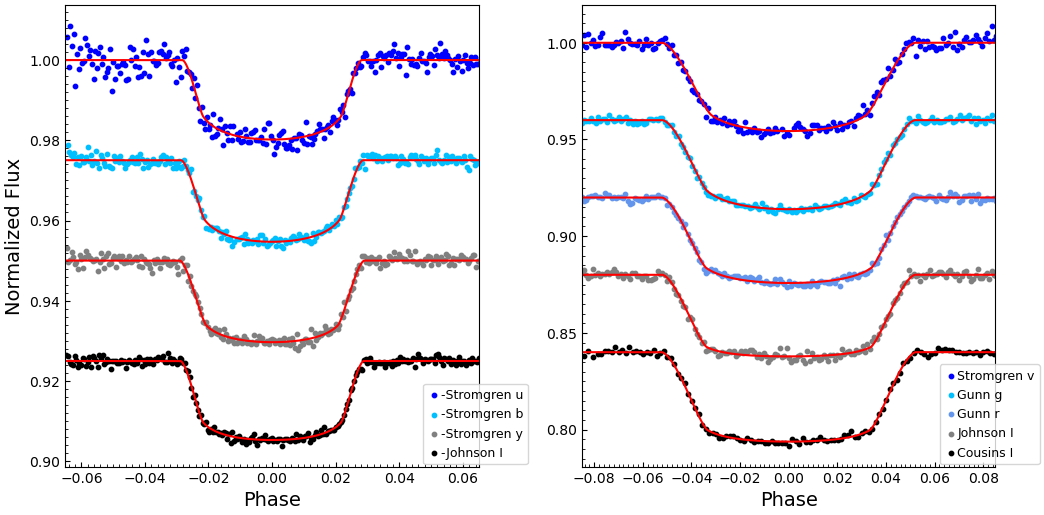}
    \caption{The ground-based light curves presented in this work for TYC 2755-36-1 (left) and TYC 3121-1659-1 (right). The filters are labelled in the legends. The red lines show the fitted models.}
    \label{ground.fig}
\end{figure*}

\begin{figure*}
    \centering
    \includegraphics[width = \textwidth]{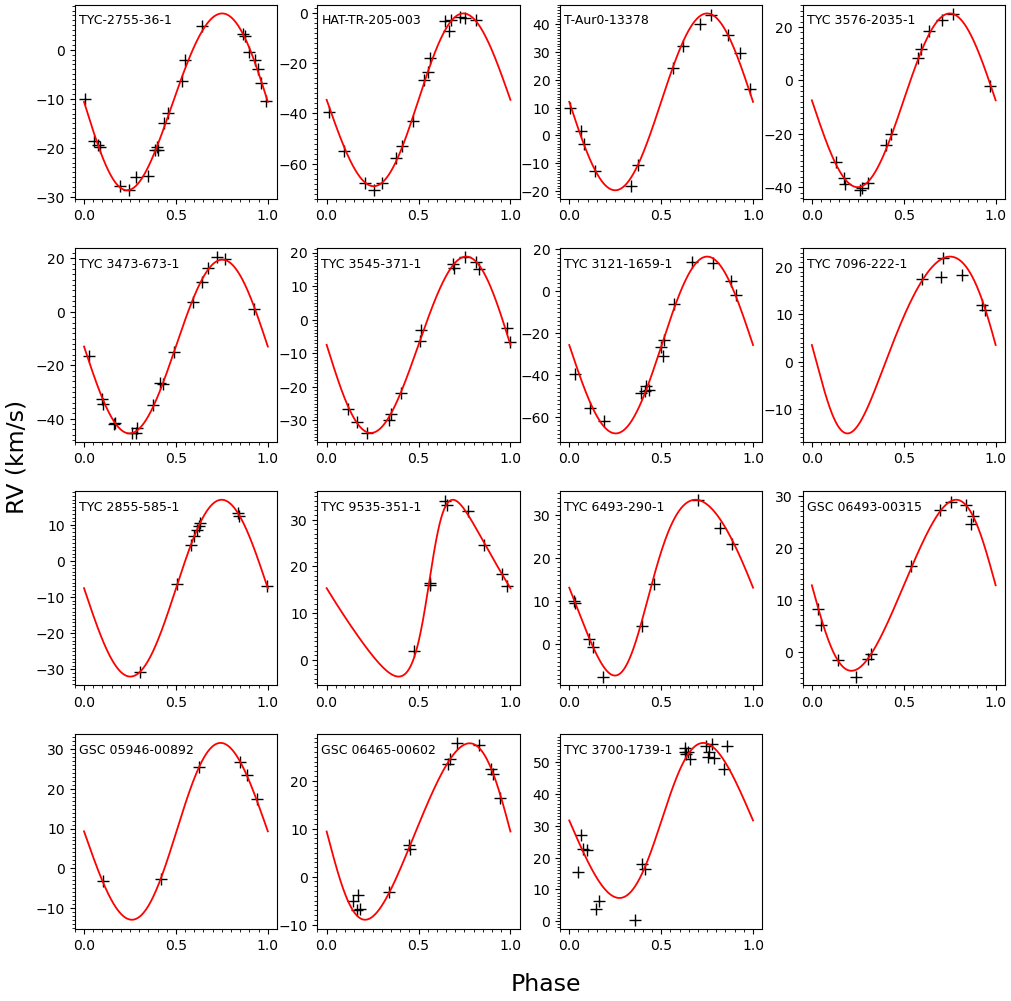}
    \caption{Fits to the RV data for each object. Object names are shown in the top left of each plot.}
    \label{RV.fig}
\end{figure*}

\subsection{Ground-based observations}

%T-Lyr0-08070: TYC 3121-1659-1
%full transit with busca in aug 2010
%full transit with caha 1.23m on 2013/06/13

%hat-tr-205-013: TYC 2755-36-1
%full transit with busca in aug 2010

One transit each of TYC 2755-36-1 and TYC 3121-1659-1 was observed simultaneously in four passbands using the Bonn University Simultaneous CAmera (BUSCA) four-band imaging photometer \citep{Reif+99spie} on the 2.2\,m telescope at Calar Alto, Spain (Fig.\,\ref{ground.fig}). Due to the brightness of TYC 2755-36-1 we elected to observe through the intermediate-band Str\"omgren $uby$ and Johnson $I$ filters. TYC 3121-1659-1 is significantly fainter and was observed through the Str\"omgren $v$, Gunn $g$ and $r$, and Johnson $I$ filters. In both cases we operated with the telescope defocussed following the approach of \citet{Me+09mn}, and were able to extract good light curves in all four passbands.

One more transit of TYC 3121-1659-1 was obtained through a Cousins $I$ filter using the 1.23\,m telescope at Calar Alto and the DLR-MKIII CCD camera. The telescope was operated out of focus as before. 

The data were reduced using the {\sc defot} pipeline \citep{Me+09mn,Me+14mn}, which depends on the NASA {\sc astrolib} library\footnote{{\tt http://idlastro.gsfc.nasa.gov/}} {\sc idl}\footnote{{\tt http://www.harrisgeospatial.com/SoftwareTechnology/ IDL.aspx}} implementation of the {\sc aper} photometry routine from {\sc daophot} \citep{Stetson87pasp}. We specified software apertures by hand and chose the aperture radii that minimised the scatter in the final light curve. Differential-magnitude light curves were generated versus an ensemble comparison star containing the weighted flux sum of all good comparison stars. A low-order polynomial was also fitted to the observations outside transit and subtracted to shift the final light curve to zero differential magnitude. The timestamps were converted to the midpoint of the exposure on the BJD$_{\rm TDB}$ timescale \citep{Eastman++10pasp}.

\section{Analysis methods}\label{Analysis.S}
\subsection{Light and RV curve modelling}

The light curves and published RVs were modelled using the \textsc{jktebop} code \citep{JKTEBOP} after converting the fluxes and errors to magnitudes and converting the time stamps from Barycentric TESS Julian Day (BTJD) to BJD. The components of the binary systems were modelled as spheres under the assumption that distortion from tidal effects would be negligible in EBLMs. We investigated the validity of this assumption by quantifying the amount of distortion expected along the lines of centres of the stars following \citet{Sterne_1941} and \citet{Beech_1985} and using the parameters derived in later sections. Assuming synchronous rotation, the average distortion expected among our list of targets is $\approx 0.36\%$. \reff{The average ratio of the uncertainty in the final radii values against the uncertainty expected from distortion is $\approx 35$ with all but one object having a ratio of at least 9. For all but one of our targets, the expected deformation is therefore insignificant compared to the size of the uncertainty on the final radii measurement. For TYC 3121-1659-1, the distortion reaches a value of $1.6\%$. This is expected given the relatively larger value of $r_1$ for this object (see Table \,\ref{lc_params.tab}).}

Parameters fitted for all targets with \textsc{jktebop} were the period $P$, the ratio of the radii $k = R_2/R_1$, the inclination $i$, the sum of the radii normalized by the semi-major axis of the orbit, $r_1+r_2$, the time of primary minimum $T_0$ and a magnitude scale factor. All targets were fitted for their surface brightness ratio $J$, except TYC 2755-36-1 for which only data for the primary eclipse is available so we assumed $J=0$. The Poincar\'e elements, $e\cos\omega$ and $e\sin\omega$, where $e$ is the orbital eccentricity and $\omega$ the argument of periastron, were also included for all targets. The quadratic limb darkening law was used to model limb darkening where as many coefficients were included in the fit as possible subject to the condition that their fitted values remained between 0 and 1. Where this condition was not satisfied, the values were taken from \citet{Claret_2017}. A parameter to account for any contaminating light sources, $L_3$, was included as a fitted parameter but was only found to be needed for TYC 2855-585-1. For the remaining targets, we found that $L_3$ had either a negative value or a value smaller than its errorbar, so we fixed it at zero. Our ability to extract measurements of $L_3$ from the light curves is limited due to the shallowness of the secondary eclipses.

\reff{We included in our analysis all available published ground-based light curves that were well sampled, covered the full primary eclipse and were obtained with a $\gtrsim1$\,m telescope. Data satisfying these criteria are available for TYC 2755-36-1, TYC 3576-2035-1, TYC 3473-673-1 and GSC 06465-00602. For the remaining objects, only the published reference time of primary minimum was included to constrain the orbital ephemeris, under the assumption of a constant orbital period}. We ensured each was converted to BJD$_{\rm TDB}$. Combining the published RV measurements within the fits allowed for the primary velocity amplitude, $K_1$, and the systemic velocity, $\gamma$, to be included as free parameters and to further constrain $P$, $T_0$, the eccentricity $e$ and argument of periastron $\omega$. RV measurements published in HJD were converted to BJD using \citet{Wright&Eastmen_2014}. For ground-based data where photometry was obtained in more than one passband simultaneously, that with the highest quality was chosen to be fitted with the RVs. 

The low sampling cadence for the LC data was accounted for using numerical integration implemented within \textsc{jktebop} \citep{Me10mn}. The model was evaluated at seven points evenly spaced within an 1800\,s interval, and the average of these points was used to compare to the observed brightness measurement.

Errors in the fitted parameters were determined via both Monte Carlo and residual permutation algorithms \citep{Me08mn}. The larger of the two errorbars was chosen for each parameter. For targets observed through more than one passband, a weighted average of the photometric parameters was taken. The resulting ephemerides and spectroscopic orbits are given in Table\,\ref{Orbital_Params.tab}, and the photometric parameters in Table\,\ref{lc_params.tab}. The {\sc jktebop} fits to the TESS data are shown in Fig.\,\ref{Tess.fig}, and for the ground-based data in Fig.\,\ref{ground.fig}. Fits to the RV measurements are shown in Fig.\,\ref{RV.fig}.

\subsection{Physical properties}\label{sec:physical_properties}

The physical properties of the systems were determined using the values of $r_1$, $r_2$, $i$, $e$, $K_1$ and $P$ found above. \reff{This analysis also used measurements of the effective temperature, \Teff, as given in Table\,\ref{basic_params.tab}, as well as published values for the metallicity, if those values were measured from high resolution spectra, e.g., at least $R\sim20000$. For cases where such a measurement is absent, we adopted $\mathrm{[Fe/H]} = -0.1 \pm 0.2$ as a representative value in the solar neighbourhood \citep[][their fig.\,3]{Haywood_2001}. Table \ref{tab: metallicity information} presents the values for [Fe/H] for objects where we used the previously published value, as well as the spectrograph and resolution used to obtain the measurement.} 

\begin{table} \caption{\label{tab: metallicity information} \reff{Previously reported values for [Fe/H] for objects where the value was measured from spectral observations with $R \gtrsim 20000$. Also given is the spectrograph and resolution used to obtain the measurement.}} 
\begin{tabular}{l r@{\,$\pm$\,}l l l }
\hline
Object & \multicolumn{2}{c}{[Fe/H]} & $R$ &  Spectrograph  \\
\hline
TYC-7096-222-1   &  0.08 & 0.13  &  60000   &    CORALIE   \\
TYC-6493-290-1   & -0.4  & 0.1   &  24000   &    ANU 2.3m Echelle \\  
GSC-06493-00315  & -0.4  & 0.1   &  24000   &    ANU 2.3m Echelle \\
GSC-05946-00892  & -0.1  & 0.2   &  24000   &    ANU 2.3m Echelle \\
GSC-06465-00602  & -0.60 & 0.06  &  24000   &    ANU 2.3m Echelle \\
TYC-3700-1739-1  & -0.05 & 0.17  &  32000   &    Tautenburg 2m Echelle \\
\hline
\end{tabular} \end{table}

Those values for \Teff\ in Table\,\ref{basic_params.tab} are taken from the corresponding reference also given in that table when the analysis is deemed reliable, such as those determined from high-resolution \'echelle spectra. In other cases, the \Teff\ values correspond to those calculated by us from a fit to the spectral energy distribution (SED), using the \textsc{vosa} software \citep{VOSA} to obtain photometric flux values as well as perform the fit. Values of the colour excess, $E(B-V)$, were determined using the \textsc{stilism} tool \citep{Lallement+18aa} and a value of $R_V = 3.1$ was used as the canonical value in the diffuse interstellar medium \citep{Schlafly_2011} resulting in extinction coefficients, $A_V$, to be included in the SED fit and allowed to adjust to within their error bar for each target. In most cases, the error was taken as half the grid step in the models that the $\chi^2$ fit was calculated against and is 125\,K in \Teff. For TYC 2755-36-1 and HAT-TR-205-003 this value was increased to 200\,K and 150\,K, respectively, in order to remain consistent with \textit{Gaia} and TESS predictions. An asterisk next to the quoted \Teff\ in Table\,\ref{basic_params.tab} indicates that the value was determined via this SED fitting process. 

For each object, we first estimated a suitable value of the velocity amplitude of the secondary component ($K_2$) and then calculated the physical properties of the system. This initial value of $K_2$ was iteratively refined to maximise the agreement between the measured \Teff \space and calculated radius of the primary star, and the predictions of theoretical models for a given mass and [Fe/H]. This was done over a grid of age values to determine the overall best mass and age for the system, and then over a set of five different sets of theoretical models \citep{Southworth_2009,Me10mn}. \reff{The model sets used were the Yonsei-Yale \citep{Demarque+04apjs}, Teramo \citep{Pietrinferni+04apj}, VRSS \citep{Vandenberg++06apjs}, Dartmouth \citep{Dotter+08apjs}, and an extension to lower masses of the models from \citet{Claret04aa}.}

Random errors from the input parameters were propagated by perturbation. Systematic errors were quantified by calculating the largest difference in values for a given parameter between the results using the five sets of theoretical models. The resulting properties are given in Table\,\ref{tab:star1} for the primary components and Table\,\ref{tab:star2} for the secondary components of our target EBLMs.

We determined the \Teff\ value of each secondary star as follows. We interpolated a synthetic spectrum for the primary star using its \Teff\ and surface gravity, and the BT-Settl synthetic spectral grid from \citet{Allard+01apj}. We then did the same for a grid of \Teff\ values for the secondary star covering 2600\,K to 6000\,K. All spectra were then convolved with the TESS passband response function from \citet{Ricker+15jatis} and the predicted surface brightness ratio within the TESS passband was calculated by dividing the secondary star spectra by the primary star spectrum. The \Teff\ of the secondary star was then obtained by interpolating the grid of surface brightness ratios to the value measured from the light curve. Errors in the measured surface brightness ratio and primary star \Teff\ were propagated and added in quadrature.

\reff{We did not make use of any \textit{Gaia} information in our analysis. This is because six of our objects have a renormalised unit weight error (RUWE) much larger than the maximum value of 1.4 for a reliable astrometric solution \citep{Gaia21aa}. However, we did cross-check our results against the distances obtained from simple inversion of the \textit{Gaia} EDR3 parallax values. To do this we adopted the physical properties determined in this work, apparent magnitudes from \citet{Hog+00aa} and \citet{2MASS}, interstellar extinction values from \citet{Lallement+18aa} and bolometric corrections from \citet{Girardi+02aa}. We found good agreement in all cases, but this is not a strong conclusion because of the significant correlation between \Teff\ and reddening, and the uncertainty of the $JHK$ apparent magnitudes and passband definitions.}

\begin{table*}\caption{The orbital ephemerides and spectroscopic orbital parameters for the objects studied in the current work. Quantities in brackets represent the uncertainties in the final digits of the preceding number.}
    \addtolength{\leftskip} {-0.5cm} % increase (absolute) value if needed
    \addtolength{\rightskip}{-0.5cm}
    \begin{adjustbox}{width=17cm}
    \centering
    \begin{tabular}{lcccccc}
    \hline 
    Object & $P$ (days) & $T_0$ (BJD$_{\rm TDB}$) & $K_1$ (km/s) & $\gamma$ (km/s)& $e$ & $\omega$ ($^\circ$)  \\
    \hline 
 
TYC 2755-36-1  &  2.230728 (13)     &  55433.49613 (13) &  18.28 $\pm$  0.38 &  -9.91 $\pm$  0.21 &  0.037 $\pm$ 0.016  &  131.5 $\pm$  26.8 \\
HAT-TR-205-003  &  2.179235216 (69) &  58744.907830 (19) &  34.48 $\pm$  0.59 &   -34.54 $\pm$  0.42 &  0.018 $\pm$  0.022 &  270.2 $\pm$   64.3 \\
T-Aur0-13378   &  3.54176081 (49) &  58820.744570 (62) &  31.78 $\pm$  0.71 &  12.09 $\pm$  0.47 &  0.000 $\pm$  0.000 &    0.0 $\pm$    0.0\\
TYC 3576-2035-1    &  6.56016357 (28) &  58702.610279 (71) &  32.46 $\pm$  0.33 &   -7.41 $\pm$ 0.27   &  0.003 $\pm$  0.004 &  107.6 $\pm$   79.0 \\
TYC 3473-673-1   &  2.539880435 (95) &  58945.441688 (45) &  32.52 $\pm$  0.38 &   -12.97 $\pm$  0.25 &  0.008 $\pm$  0.007 &   95.4 $\pm$   73.4 \\
TYC 3545-371-1   &  4.23382530 (21) &  58715.411745 (10) &  26.25 $\pm$  0.26 &   -7.43 $\pm$  0.22 &  0.023 $\pm$  0.015 &   90.3 $\pm$    1.2 \\
TYC 3121-1659-1   &  1.184756213 (71) &  58686.003373 (19) &  42.21 $\pm$  1.25 &   -25.81 $\pm$  0.73 &  0.006 $\pm$  0.019 &  321.3 $\pm$  115.7 \\
TYC 7096-222-1 &  8.95819901 (44) &  58493.788560 (84) &  18.62 $\pm$  0.82 &   5.39 $\pm$  2.97 &  0.134 $\pm$  0.021 &  138.0 $\pm$    9.7 \\
TYC 2855-585-1 &  2.40590630 (11) &  58792.98841 (11) &  24.46 $\pm$  0.10 &   -7.55 $\pm$  0.07 &  0.005 $\pm$  0.003 &  285.0 $\pm$   10.2 \\
TYC 9535-351-1 &  9.92610416 (55) &  58663.10479 (14) &  18.90 $\pm$  0.50 &  13.86 $\pm$  1.47 &  0.337 $\pm$  0.025 &  283.5 $\pm$    1.2 \\
TYC 6493-290-1 &  4.68679718 (63) &  58454.98281 (34) &  20.36 $\pm$  0.19 &  14.93 $\pm$  0.98 &  0.131 $\pm$  0.055 &  227.9 $\pm$   46.0 \\
GSC 06493-00315 &  3.636496 (13) &  58454.82048 (66) &  16.43 $\pm$  0.10 &  12.78 $\pm$  0.70 &  0.109 $\pm$  0.049 &   89.6 $\pm$    9.7 \\
GSC 05946-00892 &  3.80408653 (53) &  58483.00490 (46) &  22.24 $\pm$  0.16 &   9.20 $\pm$  0.10 &  0.028 $\pm$  0.006 &  280.2 $\pm$    6.4 \\
GSC 06465-00602 &  2.05180453 (54) &  59175.066023 (55) &  18.40 $\pm$  0.71 &   9.85 $\pm$  2.71 &  0.111 $\pm$  0.088 &  102.3 $\pm$   84.1 \\
TYC 3700-1739-1 &  1.351204584 (48) &  58800.808307 (13) &  24.41 $\pm$  1.01 &  31.61 $\pm$  1.69 &  0.069 $\pm$  0.055 &  272.8 $\pm$   92.5 \\
    \hline
    \end{tabular}
    \end{adjustbox}
    \label{Orbital_Params.tab}
\end{table*}

\begin{table*}\caption{Results of the {\sc jktebop} analysis for the objects studied in the current work.}
    \addtolength{\leftskip} {-0.5cm} % increase (absolute) value if needed
    \addtolength{\rightskip}{-0.5cm}
    \begin{adjustbox}{width=17cm}
    \centering
    \begin{tabular}{lccccccc}
    \hline 
    Object & $r_1$ & $r_2$ & Light ratio & $k$ & i $(^\circ)$ & $J$ & $L_3$ \\
    \hline 
    TYC 2755-36-1  &  0.166 $\pm$  0.002 &  0.0218 $\pm$  0.0004 &  --                    &  0.1313 $\pm$  0.0006 &  86.5 $\pm$  0.4& 0.0 fixed & 0.0 fixed\\
HAT-TR-205-003  &  0.169 $\pm$  0.007 &  0.0340 $\pm$  0.0015 &  0.001964 $\pm$  0.000423 &  0.2000 $\pm$  0.0034 &  83.2 $\pm$  0.8 &  0.056 $\pm$  0.014 &0.0 fixed\\
T-Aur0-13378   &  0.198 $\pm$  0.010 &  0.0296 $\pm$  0.0018 &  0.002154 $\pm$  0.000243 &  0.1492 $\pm$  0.0020 &  82.4 $\pm$  1.0 &  0.108 $\pm$  0.014 &0.0 fixed\\
TYC 3576-2035-1    &  0.107 $\pm$  0.0006 &  0.0233 $\pm$  0.0002 &  0.004020 $\pm$  0.000057 &  0.2182 $\pm$  0.0005 &  88.2 $\pm$  0.1 &  0.096 $\pm$  0.029 &0.0 fixed\\
TYC 3473-673-1   &  0.191 $\pm$  0.001 &  0.0330 $\pm$  0.0003 &  0.001904 $\pm$  0.000042 &  0.1730 $\pm$  0.0003 &  81.2 $\pm$  0.1 &  0.068 $\pm$  0.002 &0.0 fixed\\
TYC 3545-371-1   &  0.109 $\pm$  0.005 &  0.0216 $\pm$  0.0014 &  0.002843 $\pm$  0.000995 &  0.1984 $\pm$  0.0247 &  85.4 $\pm$  0.4 &  0.075 $\pm$  0.004 & 0.0 fixed\\
TYC 3121-1659-1   &  0.282 $\pm$  0.003 &  0.0559 $\pm$  0.0006 &  0.002276 $\pm$  0.000012 &  0.1983 $\pm$  0.0008 &  85.3 $\pm$  0.7 &  0.078 $\pm$  0.013 &0.0 fixed\\
TYC 7096-222-1 &  0.092 $\pm$  0.003 &  0.0145 $\pm$  0.0005 &  0.000936 $\pm$  0.000050 &  0.1580 $\pm$  0.0004 &  86.9 $\pm$  0.2 &  0.041 $\pm$  0.002  &0.0 fixed\\
TYC 2855-585-1 &  0.179 $\pm$  0.014 &  0.0287 $\pm$  0.0025 &  0.001639 $\pm$  0.000574 &  0.1607 $\pm$  0.0264 &  83.7 $\pm$  1.4 &  0.072 $\pm$  0.004 &0.22$\pm$0.20\\
TYC 9535-351-1 &  0.081 $\pm$  0.002 &  0.0114 $\pm$  0.0003 &  0.000838 $\pm$  0.000063 &  0.1409 $\pm$  0.0006 &  89.6 $\pm$  0.3 &  0.045 $\pm$  0.004 &0.0 fixed\\
TYC 6493-290-1 &  0.149 $\pm$  0.015 &  0.0160 $\pm$  0.0016 &  0.000872 $\pm$  0.000105 &  0.1079 $\pm$  0.0010 &  84.0 $\pm$  1.3 &  0.080 $\pm$  0.010  &0.0 fixed\\
GSC 06493-00315 &  0.130 $\pm$  0.008 &  0.0161 $\pm$  0.0010 &  0.000344 $\pm$  0.000206 &  0.1244 $\pm$  0.0010 &  89.9 $\pm$  1.9 &  0.024 $\pm$  0.014 &0.0 fixed\\
GSC 05946-00892 &  0.127 $\pm$  0.006 &  0.0175 $\pm$  0.0010 &  0.000692 $\pm$  0.000241 &  0.1379 $\pm$  0.0025 &  84.1 $\pm$  0.4 &  0.039 $\pm$  0.013 &0.0 fixed\\
GSC 06465-00602 &  0.161 $\pm$  0.005 &  0.0199 $\pm$  0.0007 &  0.000282 $\pm$  0.000564 &  0.1264 $\pm$  0.0007 &  90.0 $\pm$  1.1 &  0.012 $\pm$  0.024 &0.0 fixed\\
TYC 3700-1739-1 &  0.262 $\pm$  0.023 &  0.0468 $\pm$  0.0040 &  0.001877 $\pm$  0.000310 &  0.1784 $\pm$  0.0015 &  79.8 $\pm$  1.8 &  0.068 $\pm$  0.012 &0.0 fixed\\
    \hline
    \end{tabular}
    \end{adjustbox}
    \label{lc_params.tab}
\end{table*}

\begin{table*} \caption{\label{tab:star1} The physical properties determined in the current work for the primary stars. For these calculations we used the nominal physical constants and solar quantities defined by the IAU \citep{Prsa+16aj}. }
\begin{tabular}{l r@{\,$\pm$\,}l r@{\,$\pm$\,}l r@{\,$\pm$\,}l r@{\,$\pm$\,}l c}
\hline
Object & \multicolumn{2}{c}{$M_{\rm 1}$ (\Msun)} & \multicolumn{2}{c}{$R_{\rm 1}$    (\Rsun)} & \multicolumn{2}{c}{$\log g_{\rm 1}$ (cgs)} & \multicolumn{2}{c}{$\rho_{\rm 1}$ (\psun)} & Age (Gyr)\\
\hline
TYC 2755-36-1    & 1.241 & 0.085 & 1.312 & 0.038 & 4.296 & 0.019 & 0.549 & 0.030 & $1.6\,^{+1.5}_{-0.6}$ \\
HAT-TR-205-003   & 1.150 & 0.074 & 1.343 & 0.062 & 4.243 & 0.037 & 0.475 & 0.059 & $3.2\,^{+1.6}_{-0.7}$ \\
T-Aur0-13378     & 1.270 & 0.071 & 2.26  & 0.12  & 3.835 & 0.045 & 0.111 & 0.017 & $1.9\,^{+1.0}_{-0.4}$ \\
TYC 3576-2035-1  & 1.046 & 0.061 & 1.767 & 0.035 & 3.963 & 0.013 & 0.189 & 0.005 & $6.8\,^{+1.4}_{-1.1}$ \\
TYC 3473-673-1   & 1.161 & 0.061 & 1.686 & 0.033 & 4.050 & 0.012 & 0.242 & 0.008 & $4.5\,^{+1.2}_{-0.3}$ \\
TYC 3545-371-1   & 1.350 & 0.069 & 1.283 & 0.067 & 4.352 & 0.045 & 0.640 & 0.098 & $0.2\,^{+0.5}_{-0.2}$ \\
TYC 3121-1659-1  & 1.271 & 0.063 & 1.545 & 0.029 & 4.165 & 0.012 & 0.345 & 0.011 & $2.4\,^{+0.7}_{-0.6}$ \\
TYC 7096-222-1   & 1.67  & 0.11  & 2.087 & 0.083 & 4.022 & 0.030 & 0.184 & 0.018 & $0.5\,^{+0.4}_{-0.2}$ \\
TYC 2855-585-1   & 1.20  & 0.11  & 1.51  & 0.13  & 4.159 & 0.069 & 0.348 & 0.083 & $2.9\,^{+1.9}_{-1.1}$ \\
TYC 9535-351-1   & 1.263 & 0.072 & 1.778 & 0.055 & 4.040 & 0.024 & 0.225 & 0.017 & $2.2\,^{+0.6}_{-0.6}$ \\
TYC 6493-290-1   & 1.071 & 0.078 & 1.90  & 0.20  & 3.913 & 0.088 & 0.157 & 0.049 & $4.6\,^{+2.1}_{-0.7}$ \\
GSC 06493-00315  & 1.169 & 0.087 & 1.416 & 0.094 & 4.204 & 0.055 & 0.412 & 0.077 & $2.6\,^{+1.8}_{-0.8}$ \\
GSC 05946-00892  & 1.11  & 0.11  & 1.423 & 0.080 & 4.176 & 0.043 & 0.384 & 0.055 & $4.6\,^{+3.0}_{-0.7}$ \\
GSC 06465-00602  & 1.003 & 0.053 & 1.26  & 0.18  & 4.24  & 0.13  & 0.51  & 0.24  & $4.9\,^{+1.9}_{-1.3}$ \\
TYC 3700-1739-1  & 1.53  & 0.12  & 1.61  & 0.15  & 4.208 & 0.077 & 0.364 & 0.098 & $0.4\,^{+0.5}_{-0.2}$ \\
\hline
\end{tabular} \end{table*}

\begin{table*} \caption{\label{tab:star2} Same as Table\,\ref{tab:star1} but for the secondary components.}
\begin{tabular}{l r@{\,$\pm$\,}l r@{\,$\pm$\,}l r@{\,$\pm$\,}l r@{\,$\pm$\,}l r@{\,$\pm$\,}l r@{\,$\pm$\,}r}
\hline
Object & \multicolumn{2}{c}{$M_{\rm 2}$ (\Msun)} & \multicolumn{2}{c}{$R_{\rm 2}$ (\Rsun)} & \multicolumn{2}{c}{$\log g_{\rm 2}$ (cgs)} & \multicolumn{2}{c}{$\rho_{\rm 2}$ (\psun)} & \multicolumn{2}{c}{$a$ (\Rsun)} & \multicolumn{2}{c}{$T_{\rm eff,2}$ (K)} \\
\hline
TYC 2755-36-1   & 0.139 & 0.007 & 0.172 & 0.005 & 5.111 & 0.019 & 29.1 &  1.8 &  8.00 & 0.17 & \multicolumn{2}{c}{--} \\
HAT-TR-205-003  & 0.267 & 0.012 & 0.270 & 0.013 & 5.001 & 0.039 & 14.4 &  1.9 &  7.95 & 0.16 & 3243 & 184 \\
T-Aur0-13378    & 0.312 & 0.013 & 0.337 & 0.021 & 4.876 & 0.054 &  8.6 &  1.6 & 11.40 & 0.20 & 3808 & 128 \\
TYC 3576-2035-1 & 0.358 & 0.013 & 0.385 & 0.007 & 4.822 & 0.009 &  6.7 &  0.2 & 16.52 & 0.29 & 3440 & 235 \\
TYC 3473-673-1  & 0.267 & 0.009 & 0.291 & 0.005 & 4.937 & 0.009 & 11.5 &  0.4 &  8.83 & 0.15 & 3335 &  69 \\
TYC 3545-371-1  & 0.277 & 0.009 & 0.316 & 0.019 & 4.880 & 0.051 &  9.3 &  1.6 & 12.96 & 0.19 & 3614 &  77 \\
TYC 3121-1659-1 & 0.282 & 0.013 & 0.303 & 0.006 & 4.925 & 0.016 & 10.7 &  0.5 &  5.46 & 0.10 & 3546 & 141 \\
TYC 7096-222-1  & 0.281 & 0.018 & 0.329 & 0.013 & 4.853 & 0.036 &  8.4 &  1.0 & 22.69 & 0.50 & 3347 & 136 \\
TYC 2855-585-1  & 0.193 & 0.011 & 0.242 & 0.022 & 4.956 & 0.078 & 14.5 &  3.9 &  8.44 & 0.24 & 3449 & 137 \\
TYC 9535-351-1  & 0.231 & 0.013 & 0.251 & 0.006 & 5.001 & 0.022 & 15.5 &  1.0 & 22.22 & 0.38 & 3191 &  81 \\
TYC 6493-290-1  & 0.186 & 0.009 & 0.204 & 0.021 & 5.090 & 0.090 & 23.4 &  7.3 & 12.72 & 0.28 & 3483 & 126 \\
GSC 06493-00315 & 0.141 & 0.007 & 0.175 & 0.012 & 5.100 & 0.055 & 27.9 &  5.3 & 10.89 & 0.25 & 2859 & 343 \\
GSC 05946-00892 & 0.196 & 0.012 & 0.196 & 0.013 & 5.145 & 0.050 & 27.6 &  4.8 & 11.21 & 0.33 & 3003 & 226 \\
GSC 06465-00602 & 0.118 & 0.006 & 0.199 & 0.029 & 4.911 & 0.137 & 15.9 &  7.4 &  7.06 & 0.15 & 2551 & 593 \\
TYC 3700-1739-1 & 0.184 & 0.012 & 0.288 & 0.026 & 4.784 & 0.078 &  8.2 &  2.2 &  6.16 & 0.16 & 3646 & 190 \\
\hline
\end{tabular} \end{table*}

\section{Results for each system}\label{sec:results}\label{Results.S}

\subsection{TYC 2755-36-1}

TYC 2755-36-1 was observed by the HATNet (Hungarian-made Automatic Telescope Network) wide angle survey and identified as a planetary candidate by both \citet{Beatty_2007} and \citet{Latham_2009}. Both studies reported the object to be an EB with a faint component following reconnaissance spectroscopy. The 23 RV measurements published by \citet{Latham_2009} are identical to those obtained by \citet{Beatty_2007}. These measurements are presented in their table~1 and were utilized in the current study (Fig.\,\ref{RV.fig}). %As stated previously, all RV measurements obtained by previous authors of the objects included in this work, and discussed in the following subsections, are those which were used in the current analysis and shown in Fig.\,\ref{RV.fig}. 

Stellar parameters were also reported by both \citet{Beatty_2007} and \citet{Latham_2009} for TYC 2755-36-1 via cross correlation of the observed spectra against a library of synthetic spectra. A full characterization of the system was performed by \citet{Beatty_2007} using the assumption that the system is synchronized and combining their spectroscopic measurements with those derived from the modelling of follow-up light curves. The resulting masses and radii from their study can be found in Table\,\ref{tab:literature} along with the previous literature estimations of these parameters for all other objects included in this work.

\reff{The radii reported by \citet{Beatty_2007} for both components of TYC 2755-36-1 agree with our estimations within $1\sigma$}. The masses are larger by $19\%$ ($1.2\sigma$) and $12\%$ ($1.2\sigma$) for the primary and secondary, respectively. The previous authors report an M~dwarf radius inflated by $11\%$ while the current study finds that the magnitude of inflation is $5\%$ when compared to the corresponding \citet{Baraffe_2015} (BCAH15, hereafter) isochrone for its age estimation given in Table\,\ref{tab:star1}. %The error bar for the age estimation of this object spans the pre-main-sequence; it is therefore unclear how reliably the inflation status of this star can be determined given the size of the variation in the predicted radius over the lower range of its age uncertainty. This is shown by comparing the position of the 0.05 Gyr and 0.3 Gyr isochrones in Fig.\,\ref{fig:mass-radius}.

\reff{The M~dwarf \reff{in this system} is the densest and has the smallest radius out of all objects included in the current study. It has the second largest surface gravity and is the second least massive. 
%The high density would oppose the idea that its true age could be located near the lower end of its error bar on the pre-main-sequence and so suggests the absolute age presented in Table\,\ref{tab:star1} is well estimated. Then, assuming the inflation status of the M~dwarf by both studies is accurate, 
There are no spots identified in the light curve of the host system, and this would oppose the hypothesis that enhanced magnetic fields due to faster rotation induced by synchronization are the cause for inflation \citep{Beatty_2007}.}

\begin{table*} \caption{\label{tab:literature} Results from previous authors. For TYC 7096-222-1, the quoted value for $M_2$ is in the middle of the range off possible values reported by the previous authors and the given uncertainty satisfies both ends of this range.}
\begin{tabular}{l r@{\,$\pm$\,}l r@{\,$\pm$\,}l r@{\,$\pm$\,}l r@{\,$\pm$\,}l}
\hline
Object & \multicolumn{2}{c}{$M_1$ (\Msun)} & \multicolumn{2}{c}{$M_2$ (\Msun)} & \multicolumn{2}{c}{$R_1$ (\Rsun)} & \multicolumn{2}{c}{$R_2$ (\Rsun)} \\
\hline
TYC 2755-36-1   & 1.04  & 0.14  & 0.124 & 0.011 & 1.28  & 0.04  & 0.169 & 0.006   \\
T-Aur0-13378    & 1.60  & 0.13  & 0.37  & 0.03  & 2.40  & 0.10  & 0.37  & 0.02    \\
TYC 3576-2035-1 & 0.91  & 0.15  & 0.345 & 0.034 & 1.63  & 0.08  & 0.360 & 0.017   \\
TYC 3473-673-1  & 1.49  & 0.07  & 0.315 & 0.010 & 1.83  & 0.03  & 0.325 & 0.005   \\
TYC 3545-371-1  & 0.77  & 0.08  & 0.198 & 0.012 & 1.14  & 0.03  & 0.238 & 0.007   \\
TYC 3121-1659-1 & 0.95  & 0.11  & 0.240 & 0.019 & 1.36  & 0.05  & 0.265 & 0.010   \\
TYC 7096-222-1  & 1.735 & 0.054 & 0.435 & 0.165 & 1.662 & 0.092 &  0.263  & 0.046    \\
TYC 2855-585-1  & 1.26  & 0.11 & 0.20  & 0.02  & 1.30  & 0.05  & 0.17  & 0.01    \\
TYC 6493-290-1  & 1.10  & 0.07  & 0.17  & 0.01  & 1.70  & 0.09  & 0.18  & 0.01    \\
GSC 06493-00315 & 1.1   & 0.1   & 0.132 & 0.01  & 1.20  & 0.05  & 0.154 & 0.007   \\
GSC 05946-00892 & 1.2   & 0.1   & 0.20  & 0.02  & 1.58  & 0.06  & 0.22  & 0.01    \\
GSC 06465-00602 & 0.97  & 0.06 & 0.110 & 0.006 & 1.22  & 0.03  & 0.147 & 0.004   \\
TYC 3700-1739-1 & 1.493 & 0.073 & 0.188 & 0.014 & 1.474 & 0.040 & 0.234 & 0.009   \\
\hline
\end{tabular} \end{table*}

\subsection{HAT-TR-205-003}

HAT-TR-205-003 was also observed by the HATNet survey and identified as a planetary transit candidate by \citet{Latham_2009}. A spectroscopic reconnaissance confirmed the nature of the companion to be stellar. Seventeen RV measurements were obtained from single-order \'echelle spectra and used to obtain a preliminary orbital solution. These RV measurements are presented in table~2 of their paper. 

By cross-correlating the observed spectra against a library of synthetic spectra, \citet{Latham_2009} were able to derive estimates for the \Teff, $v \sin i$ and $\log(g)$ assuming solar metallicity, but aside from a preliminary estimation of $R_1$, no further characterization of the object was carried out as their study concentrated on candidate transiting planets. 

Our results are the first measurements of the fundamental parameters for this object. The radius of the companion in this system, $R_2 = 0.270 \pm 0.013 \Rsun$, is found to agree with the BCAH15 theoretical predictions for its mass ($M_2 = 0.267 \pm 0.012\Msun$) and age ($3.2^{+1.6}_{-0.7}$ Gyr).

\subsection{T-Aur0-13378}\label{T-Aur0-13378}

The transit events for T-Aur0-13378, as well as the objects described in the following four subsections, were detected by the Trans-Atlantic Exoplanet Survey (TrES) and the companion in each was confirmed to be stellar by \citet{Fernandez_2009} via a spectroscopic reconnaissance. Thirteen RVs were measured for each object from single-order \'echelle spectra along with estimations for \Teff, $v \sin i$ and $\log(g)$. The RVs are presented in their tables 2--6. The stellar parameters were derived for four fixed metallicities, $-1$, $-0.5$, 0.0 and $+0.5$ dex, resulting in four sets of parameters. Binary parameters were obtained from the analysis of follow-up light curves.

Combining binary and spectroscopic parameters allowed for the derivation of physical parameters via isochrone fitting following the procedure of \citet{Torres_2008}. Physical parameters were also derived independently by invoking the assumption that the orbits are synchronized. Both methods were carried out for each set of atmospheric parameters corresponding to the four fixed metallicities described above. The expectation was that one of the fixed metallicities would yield agreeable solutions between the two methods. 

For T-Aur0-13378, no good solution was found for any of the input metallicities, opposing the idea that the system is synchronized, so the results from isochrone fitting with solar abundance were adopted. It was necessary to fix $e=0$ in our study to obtain the optimum solution and this indicates that the system is at least older than the circularization timescale. However, \citet{Fernandez_2009} calculated the ratio of the orbital and rotational angular momentum, $\alpha$, for the objects in their study and for each of them, this resulted in a value larger than 70, which means that the timescales for synchronization and alignment are expected to be shorter than that for circularization. 

\citet{Fernandez_2009} hypothesize a possible reason for the system not being synchronized but circularized is that the primary component in the system is evolved. As the primary component expands, its rotational velocity decreases in order to conserve angular momentum and this process may dominate over tidal forces acting to synchronize the system. The current results support this hypothesis with small values for the surface gravity and density for a relatively massive star among Table\,\ref{tab:star1}, along with the relatively large surface brightness ratio reported in Table\,\ref{lc_params.tab}.

\reff{The masses derived here are smaller than those reported by the previous authors, and given in Table\,\ref{tab:literature}, by $21\%$ ($2.2\sigma$) and $16\%$ ($1.8\sigma$) for the primary and secondary components, respectively. The estimated radius for the primary component is consistent between both studies within $1\sigma$ but our measurement of the M~dwarf radius is $9\%$ ($1.1\sigma$) smaller.} A marginal inconsistency in the radius of the M~dwarf was observed by \citet{Fernandez_2009} when compared to the theory in the direction of inflation. The current results demonstrate inflation by $12\%$ relative to the BCAH15 isochrone for its age.

\subsection{TYC 3576-2035-1}

The study by \citet{Fernandez_2009} also included the system TYC 3576-2035-1. For this system, there was a metallicity range where the two methods of characterization agreed and a value of [Fe/H] $=-0.5 \pm 0.2$ dex was adopted. For systems where an acceptable solution was found, \citet{Fernandez_2009} adopted the one derived via the assumption of synchronization.

\reff{The primary star and M~dwarf masses from that solution are both in agreement with the parameters derived here within $1\sigma$. Our estimations for the radii are $8\%$ ($1.6\sigma$) and $7\%$ ($1.4\sigma$) larger compared to those of \citet{Fernandez_2009} for $R_1$ and $R_2$, respectively.} The previous authors find their estimation for the radius of the M~dwarf to be marginally inflated compared to isochrones. We observe a radius $11\%$ larger than the BCAH15 theoretical prediction for its age. 

The secondary eclipse of about 0.0025 mag deep is clearly visible and, combined with the good phase coverage in the RV measurements, provides a strong constraint on the eccentricity for this system of $e = 0.003 \pm 0.004$ (Table\,\ref{Orbital_Params.tab}). This is consistent with a circular orbit and shows that circularization timescales have been satisfied by the age of this system. Given the value of $\alpha>70$ reported by \citet{Fernandez_2009}, the system is expected to be synchronized.

As \citet{Fernandez_2009} state, synchronization is not always guaranteed when evidence is strong (e.g, \citealt{Pont_2006}) and the consistency between their synchronized and model-dependent solutions is ultimately dependent on the adopted atmospheric parameters. \reff{The disagreement between our results for the radii and those of the synchronized solution reported by \citet{Fernandez_2009} may therefore be explained by the difference in metal abundance and \Teff\ adopted between the studies.}

\subsection{TYC 3473-673-1}

For TYC 3473-673-1, \citet{Fernandez_2009} deduced a metallicity of [Fe/H] $= 0.5 \pm 0.2$ dex along with the parameters corresponding to a synchronized orbit. The primary and secondary masses derived here are $22\%$ ($3.5\sigma$) and $15\%$ ($3.6\sigma$) smaller compared to the values derived by \citet{Fernandez_2009} while the radii are $8\%$ ($3.2\sigma$) and $10\%$ ($4.8\sigma$) smaller. 

Again, good phase coverage in RV measurements and a visible secondary eclipse leads to a well-constrained orbit for this system with $e = 0.008 \pm 0.007$. This is consistent with a circular orbit \citep{LucySweeney_1971}, so the assumption of synchronization invoked by \citet{Fernandez_2009} is valid given $\alpha>70$. The arguments explained in the previous subsection apply and may explain the disagreement in the resulting parameters of the two studies. The finding that the M~dwarf is inflated compared to models is, however, in agreement with our findings, where we find that its radius is $7\%$ larger than the BCAH15 predictions.

\subsection{TYC 3545-371-1}

A solution corresponding to a synchronized orbit was adopted by \citet{Fernandez_2009} for TYC 3545-371-1 on the basis that the resulting parameters agreed with those from isochrone fitting for a metallicity of [Fe/H] $=-0.5 \pm 0.2$ dex. Our results are larger by $75\%$ ($5.5\sigma$) and $40\%$ ($5.3\sigma$) in mass and $13\%$ ($2.0\sigma$) and $33\%$ ($3.9\sigma$) in radius for the primary and secondary, respectively. These are large discrepancies and are probably caused by differences in the assumed \Teff\ values between the two studies. We adopted a \Teff\ value 756\,K hotter than that of the model-dependent solution derived by \citet{Fernandez_2009}, which matched the synchronized solution that they adopted. Also, the difference in metal abundance is 0.5\,dex. 

\reff{Their result for $R_2$ was found to disagree within error with theoretical predictions in the direction of radius inflation. The prediction here is that the M~dwarf radius is inflated compared to theory for a pre-MS age of $\sim$0.2\,Gyr (reported in Table\,\ref{tab:star1}). The error bar for the age estimation of this object, however, spans the entire pre-main-sequence; it is therefore unclear how reliably the inflation status of this star can be determined given the size of the variation in the predicted radius over the lower range of its age uncertainty. This is shown by comparing the position of the 0.05 Gyr and 0.3 Gyr isochrones in Fig.\,\ref{fig:mass-radius}.}
%In the mass range of the secondary component, Fig.\,\ref{fig:mass-radius} shows that differences between theoretical radii predictions corresponding to 0.3\,Gyr and 10\,Gyr are negligible. 

A significant value for third light of \emph{$L_3 = 0.374 \pm 0.127$} was found in our first analysis. However, this leads to an unphysically small $r_1$ so we instead present a solution with $L_3 = 0$. The \Teff\ we use has a noticeable effect on the age estimate for this target -- a lower \Teff\ of 6500\,K gives a solution with a larger age of $1.5^{+0.5}_{-0.6}$\,Gyr and a significantly lower $M_1$ of $1.292 \pm 0.055$\Msun. A high-quality spectrum of this target would be useful for checking and confirming its \Teff.

\subsection{TYC 3121-1659-1}

The fifth and final object studied by \citet{Fernandez_2009} is TYC 3121-1659-1. A synchronized solution, matched to a model-dependent solution with [Fe/H] $= -0.5 \pm 0.2$ dex, was adopted. Our estimates of the fundamental parameters for this object were again in conflict with those derived by \citet{Fernandez_2009}. For the primary and secondary, this amounted to a $34\%$ ($2.5\sigma$) and $18\%$ ($1.8\sigma$) increase in mass accompanied by a $14\%$ ($3.2\sigma$) and $14\%$ ($3.3\sigma$) increase in radius. The previous authors observed the M~dwarf radius to be marginally inflated, in agreement with our findings that its radius is $8.2\%$ inflated compared to the BCAH15 models. The results obtained by \citet{Fernandez_2009} are given in Table\,\ref{tab:literature} for comparison. \reff{We also calculated the properties of the system after accounting for the distortion of the primary star and found that they changed by much less than their errorbars. We therefore elected to present the results obtained without accounting for distortion, for consistency with the other objects in the current work.}

For this object, our ground-based light curves (Section \ref{Observations.S}) were used in the photometric analysis. Table\,\ref{tab:tlyr0} shows the results from the different light curves for the photometric parameters. The resulting values for $i$ and $k$ are significantly smaller from the TESS band compared to the other passbands. This may be due to the combined effect of the under-sampling of positions of contact across the primary eclipse from the TESS 30-minute cadence mode, as well as some third light being collected in the larger TESS pixels. An alternative explanation for the inclination being different is that the orbital plane has undergone precession due to exterior forcing, such as a third body. However, since the radius ratio is also affected, the former explanation is more likely. 

\begin{table*} \caption{\label{tab:tlyr0} Comparison of the results obtained for TYC 3121-1659-1.} 
\begin{tabular}{l r@{\,$\pm$\,}l r@{\,$\pm$\,}l r@{\,$\pm$\,}l r@{\,$\pm$\,}l}
\hline
Passband & \multicolumn{2}{c}{$r_1$} & \multicolumn{2}{c}{$r_2$} & \multicolumn{2}{c}{$k$} & \multicolumn{2}{c}{$i$ ($^\circ$)}  \\
\hline
TESS & 0.296 & 0.018 & 0.0522 & 0.0040 & 0.176 & 0.003 & 81.8 & 2.6 \\
Cousins $I$ & 0.280&0.004&0.0571&0.0012&0.204&0.001&86.9&2.4 \\
Gunn $r$ & 0.283&0.005&0.0554&0.0012&0.196&0.002&86.5&1.8 \\
Gunn $g$ & 0.282&0.005&0.0557&0.0013&0.198&0.002&86.2&1.6 \\
Johnson $i $& 0.281&0.005&0.0553&0.0011&0.197&0.002&86.5&1.7 \\
Str\"omgren $v$ & 0.290&0.008&0.0578&0.0020&0.199&0.003&83.5&1.5 \\
Adopted & 0.282 & 0.003 & 0.0559 &0.0006 & 0.198 & 0.001 & 85.3 & 0.7\\
\hline
\end{tabular} \end{table*}

%{\bf JKT: this whole paragraph feels very repetitive. Most or all could be cut. Just include any useful points not made anywhere else.}
%\citet{Fernandez_2009} characterized the object discussed in this sub section along with those discussed in the previous four. Agreement between the two studies was weak, with no agreement between any mass or radius estimation for TYC 3473-673-1, TYC 3545-371-1 and TYC 3121-1659-1. With the exception of TYC 3545-371-1, the discrepancies are not large when considering the $\sigma$ values in most cases. It also is possible that the disagreement is influenced by an over estimate in the precision of either set of results. 

\subsection{TYC 7096-222-1}

This object was identified as an EB using photometry collected by WASP-South and was characterized by \citet{Bentley_2009}. RVs were derived from grating spectra, and stellar parameters of the primary were determined from synthetic spectral fits to high-resolution \'echelle spectra ($R \sim 60000$). Their value for [Fe/H] of $0.08 \pm 0.13$ was adopted in the current work.     

By combining the results from their spectral and light curve analysis, \citet{Bentley_2009} derived full system parameters following the results of the primary mass, which was obtained from isochrone fitting. The mass of the secondary was found to depend on the eccentricity, which was not well constrained by the available RVs. \citet{Bentley_2009} obtained limiting values of $0.29 \pm 0.02$\Msun\ for $e = 0$, and $0.54 \pm 0.06$\Msun\ for $e = 0.75$.

Our analysis is able to rectify this issue because the secondary eclipse is detectable in the TESS observations. We find a significant eccentricity of $e = 0.134 \pm 0.021$. Our estimate of $M_2 = 0.281 \pm 0.018$\Msun\ agrees with the value quoted by \citet{Bentley_2009} for an orbit of zero or modest eccentricity while our estimate for $M_1$ is \reff{also in agreement within $1\sigma$}. The primary and secondary radii derived here are both $25\%$ larger than those calculated by \citet{Bentley_2009}, corresponding to $3.4\sigma$ and $1.4\sigma$, respectively.

\citet{Bentley_2009} find their estimation for the M~dwarf radius to agree with theoretical predictions. We find that the system is inflated by $19.3\%$, which is comparable to the $25\%$ increase in our $R_2$ \reff{compared to} \citeauthor{Bentley_2009}'s value, explaining the different conclusions regarding the object's inflation status.

The primary star is of approximately A8/F0 spectral type and such stars are expected to have rotational velocities of $\approx 200$\,km\,s$^{-1}$ \citep{gray_2005}. \citet{Bentley_2009} suggest that the object is a fairly typical Am star given its relatively slow rotation ($v \sin i = 35 \pm 0.5$\,km\,s$^{-1}$); Am stars are thought to have been spun down by a companion. The previous authors determine that the rotation of the primary is still larger than that of the secondary, showing that the synchronization process is ongoing, suggesting an upper limit to the age of the system of $\approx0.92$~Gyr, calculated as the synchronization timescale by \citet{Bentley_2009}. The current age estimation of $0.5^{+0.4}_{-0.2}$ Gyr supports these statements. \reff{It is also noted that the newly derived estimate for the eccentricity shows that circularization time-scales are also yet to be satisfied. This also supports our young age estimate for this system}.

\subsection{TYC 2855-585-1}

\citet{Koo_2012} identified transit-like variations in the photometric data of \citet{Lee_2008} for the object TYC 2855-585-1. High-resolution multi-epoch \'echelle spectra were obtained from which six RVs were derived. Follow-up photometry was modelled and the absolute dimensions were computed by applying the photometric and spectroscopic measurements to the mass--radius and mass--\Teff\ relations for EBs from \citet{Southworth_2009}.

The current estimations for the masses of the components agree with those by \citet{Koo_2012} within $1\sigma$ but the radii are $16\%$ ($1.5\sigma$) and $42\%$ ($3.0\sigma$) larger for the primary and secondary, respectively. There was no discussion of the system parameters in the context of the radius discrepancy in low mass stars by \citet{Koo_2012}, but here we find that the radius is $14.6\%$ inflated compared to its corresponding BCAH15 isochrone. 

This object is the only system where we did not fix third light at zero. We find a value of $L_3 = 0.22 \pm 0.20$, which is not significant and has little effect on the results.

\subsection{TYC 9535-351-1}

TYC 9535-351-1 was identified by \citet{Crouzet_2013} and \citet{Wang_2014} as a potential planetary candidate. Follow-up RV observations were conducted by both authors and the RVs derived confirmed the object to be an EB in both cases. Five RV measurements were provided to us by \citet{Wang_2014} and four were obtained from \citet{Crouzet_2013}.

The current study is the first time that a full characterization has been performed for the object and as such contributes another M~dwarf with precise mass and radius measurements to the literature. The orbit for this system is the most eccentric in this study, with $e=0.337 \pm 0.025$. We find that the system is $4.3\%$ inflated compared to the BCAH15 isochrone for its mass and age given in Tables \ref{tab:star2} and \ref{tab:star1}, respectively.

\subsection{TYC 6493-290-1}

TYC 6493-290-1 was identified from photometric observations by the HATSouth global network and characterized by \citet{Zhou_2014}. Spectroscopic analysis yielded ten RVs as well as atmospheric parameters of the primary. A global fit to the RVs and available light curves was performed simultaneously. The masses and radii of the components were derived at each iteration by combining the assumption of synchronization with an isochrone fitting method. For TYC 6493-290-1, their photometric follow-up data consisted of only a partial primary eclipse and so is heavily reliant on the discovery data. 

\reff{Our estimation for the primary mass agrees with the estimation from \citet{Zhou_2014} within $1\sigma$ but our value for the M~dwarf mass is $9\%$ ($1.2\sigma$) larger. The radii predictions between the two studies for both components in this system are in agreement within $1\sigma$. However,} a comparison of our values to those of \citet{Zhou_2014} is limited by the precision of the published values, so the discussion of the agreement in radii is based on rounding ours to the same precision. \citet{Zhou_2014} find the radius of the M~dwarf to agree with theoretical predictions, as do we. We find the system to be eccentric, $e = 0.131 \pm 0.055$, with an estimated age of $\approx 4.6$  Gyr.

\subsection{GSC 06493-00315}

The study by \citet{Zhou_2014} also included the system GSC 06493-00315. The identification and characterization of the object were carried out in the same way as for TYC 6493-290-1. For this object, they obtained 12 RV measurements but no follow-up light curves. The characterization was thus reliant on the discovery data alone. 

\reff{Our results for the masses of the components are in agreement with those of \citet{Zhou_2014} within $1\sigma$ but our radius measurements are $18\%$ $(2\sigma)$ and $14\%$ $(1.5\sigma)$ larger for the primary and secondary, respectively.} Again, we find a marginally significant eccentricity, $e = 0.109 \pm 0.049$, which means the assumption of synchronization \citep{Zhou_2014} is questionable. \citet{Zhou_2014} found their measured $R_2$ to agree with theoretical models within the uncertainty. Our results confirm this.

\subsection{GSC 05946-00892}

GSC 05946-00892 is another object that was previously characterized in the study by \citet{Zhou_2014}, who obtained six RV measurements. Excellent agreement is found between their estimations and our measurements of the masses of the components, amounting to less than $0.3\sigma$ for the secondary. Our radii are smaller by $10\%$ $(1.6\sigma)$ and $11\%$ $(1.5\sigma)$. \citet{Zhou_2014} derived and adopted a chemical abundance consistent with solar, [Fe/H] $=-0.1 \pm 0.2$ dex, which is in agreement with our approach.

\citet{Zhou_2014} found that $R_2$ agrees with theoretical models to within its errorbar. We find that the $R_2$ value is $8.6\%$ lower than the predicted value from the BCAH15 isochrones. Whilst the error bars of the $R_2$ values from each study encompass the other, the sum of the uncertainties is $\approx 11\%$ so is larger than the size of the under-inflation observed in the current study.

\subsection{GSC 06465-00602}

The fourth and final object characterized by \citet{Zhou_2014} is GSC 06465-00602, for which they obtained 14 RV measurements. Our measurement for the M~dwarf radius is the only one that differs with that found by \citet{Zhou_2014}, and by an amount of $35\%$ ($1.8\sigma$); all other values are in agreement within $1\sigma$. 
\citet{Zhou_2014} found that their radius estimation for the secondary is inflated by $\approx 13\%$ when compared to theoretical models. We find that the M~dwarf is inflated by $39\%$. Looking at Fig.\,\ref{fig:mass-radius}, the M~dwarf is seen to lie on the 0.1\,Gyr isochrone. 

\reff{The reported age in Table \ref{tab:star2} is estimated as $4.9^{+1.9}_{-1.3}$\,Gyr on the MS but an alternative solution that we derived, assuming solar metallicity, yields an age estimation with a lower boundary of the uncertainty that spans the pre-MS, where contraction is ongoing. This scenario would suggest that the location of the object in Fig.\,\ref{fig:mass-radius} relative to the 0.1\,Gyr isochrone might be correct, in which case the M~dwarf is not inflated. This may be important given the unusually low metallicity of [Fe/H] $= -0.6 \pm 0.06$ reported by \citet{Zhoe_2015} that was also adopted in our final solution.}
%The inflation status of this object is therefore unclear without a more precise age estimation, regardless of the size of the previously quoted value of $32.6\%$. 

The secondary component of this system has the smallest mass out of all objects included in the current study as well as the smallest surface brightness ratio. We find a marginally significant eccentricity for this system of $e = 0.111 \pm 0.088$, so the assumption of synchronization invoked by \citet{Zhou_2014} may not be valid. 

%Comparing the results here with those by \citet{Zhou_2014}, we note that the analysis by \citet{Zhou_2014} for TYC 6493-290-1 is heavily reliant on the discovery data while that for GSC 06493-00315 is fully reliant.

\subsection{TYC 3700-1739-1}

TYC 3700-1739-1 was detected in the Berlin Exoplanet Search Telescope (BEST) and Tautenburg Exoplanet Search Telescope (TEST) surveys as an exoplanetary candidate and the object was published as an uncharacterized Algol type by \citet{Pasternacki_2011} after its planetary status was deemed false. \citet{Eigmuller_2016} later combined the data from both surveys in their study of the binary star system. Spectroscopic follow-up observations were performed resulting in 21 \'echelle spectra and RV measurements. A simultaneous fit to the RVs and photometric data from the two surveys was performed and the resulting binary parameters are presented in their table~4. By co-adding the individual observations corrected for their orbital motion, they were able to obtain a high-S/N spectrum which was used to derive the stellar parameters of the primary component. Full system parameters were derived via isochrones in combination with the stellar parameters of the primary and 2MASS apparent magnitudes. The resulting parameters are in agreement with our findings to within $1\sigma$ except for $R_2$, where our estimation is $23\%$ larger, corresponding to $2.0\sigma$.

Comparisons to theoretical models by the previous authors indicate that the secondary has an inflated radius even when correcting the isochrones for 5\% of the discrepancy observed among low-mass stars. This might be due to the fact that the system is young; \citet{Eigmuller_2016} estimate a synchronization factor of $P_{\rm rot}/P_{\rm orb} = 0.43 \pm 0.05$ leading to an upper limit to the age of this system of 120--250~Myr via synchronization timescales. Those authors also derive a statistically insignificant value for the eccentricity, $e = 0.070 \pm 0.063$, in agreement with our value (Table\,\ref{Orbital_Params.tab}). This suggests that the system is at least older than the time taken to circularize and, coupled with the upper limit suggested above, tightly constrains its age.   

This upper limit \reff{on the age of the system} is encompassed by the uncertainty in the current age estimation of $0.4\,^{+0.5}_{-0.2}$. The current study estimates that the M~dwarf is $41\%$ inflated compared to a 0.3 Gyr BCAH15 isochrone. This value is suspiciously large. Looking at Fig.\,\ref{fig:mass-radius}, we see that the BCAH15 isochrones predict a rapid phase of contraction from 0.05 Gyr to 0.3 Gyr. Lowering the current age estimation within its uncertainty would account for much of the $41\%$ inflation as well as obey the upper age limit suggested above. Then, given the size of the uncertainty on the M~dwarf's age about a phase of such rapid contraction means that the inflation status of this object can not be reliably discerned, while the evidence does suggest a younger age and a lesser amount of inflation. Further support that the system is young is also provided by the overestimate of the M~dwarf \Teff\ compared to the models (see Fig.\,\ref{fig:mass-teff}), \reff{since} radius inflation is commonly observed together with a lower \Teff.

\section{Discussion}\label{Discussion.S}

We have characterized 15 EBLM stars using new light curves and published RVs. We used light curves from the TESS mission for 14 of them, and our own high-precision multi-band ground-based light curves for two of them. Two of the EBLMs had not previously been analysed in detail so our results represent the first measurements of their physical properties. Regarding those that were previously characterized, the precision in 19 of the 26 masses has been improved. \reff{Although the overall precision in the radius measurements from both studies were comparable} ($\sim 3 \%$), only 8 of the 26 radii have improved in precision, despite the much better data available, which suggests that some \reff{of the} published errorbars are underestimated. 

In order to convert the light curve and RV results into the physical parameters of the two stars, we needed a value for the metal abundance of each system. \reff{We have assumed a value representative of the solar neighbourhood in all cases as these are nearby stars, except where the previously published value was measured from high resolution spectroscopy. In the latter case, we deemed the result reliable and adopted its value. Metallicity values were published for some of the other objects, but were either derived from low-resolution spectra or have large uncertainties.}
%The exceptions to this are GSC 06465-00602, TYC 6493-290-1 and GSC 06493-00315, where \citet{Zhou_2014} report metallicities of [Fe/H] $= -0.60 \pm 0.06$, $-0.4 \pm 0.1$ and $-0.4 \pm 0.1 $ dex, respectively. These metallicities are surprisingly low and were obtained from spectra with relatively low signal-to-noise (S/N). 
\begin{figure}
    \centering
    \includegraphics[width = 8.5cm]{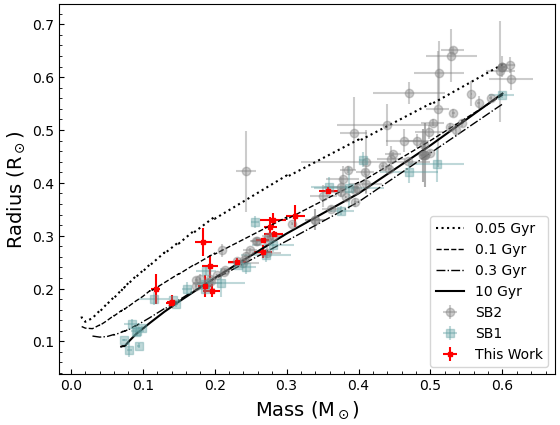}
    \caption{BCAH15 isochrones for 0.05, 0.1, 0.3 and 10 Gyr plotted in the mass-radius plane with the locations of the current objects overplotted in red. M~dwarfs from \citet{Parsons_2018} are also shown where SB2 M-dwarfs are plotted in grey and SB1 M-dwarfs are plotted in blue.}
    \label{fig:mass-radius}
\end{figure}
\begin{figure}
    \centering
    \includegraphics[width = 8.5cm]{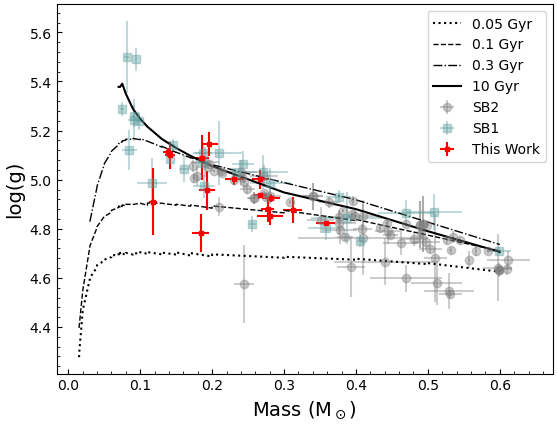}
    \caption{Same as Fig.\ref{fig:mass-radius} in the mass-log(g) plane.}
    \label{fig:mass-log(g)}
\end{figure}

Revised metallicity measurements from high-resolution and high-S/N \'echelle spectra would be valuable in most cases. Inaccuracies in the adopted metallicity leads to inaccuracies in the stellar models used in determining the physical parameters, and adds further uncertainty to the systematic uncertainty already inherent to the model-dependent results. See \citet{Southworth_2009} for a discussion of the nature in which systematic uncertainties propagate into the resulting parameters. Reliable estimations for \Teff\ could also be derived from such spectra. While the choice of metallicity affects the model used in each iteration, the choice of \Teff\ affects the best-matching solution that is returned because the solution is compared to the \Teff\ (and calculated radius) at each iteration. It is therefore important that this value is accurate such that the final best-fitting match is indeed that which corresponds to the true system parameters. 

We find that in the absence of such measurements for \Teff\ and [Fe/H], the  detection of the secondary eclipse as well as the use of high-precision space photometry still allows for significantly increased accuracy and reliability of the measured properties. This is especially the case for objects whose previous estimates for eccentricity were vague, e.g.\ TYC 7096-222-1. The literature value of [Fe/H] for this system, which we adopted, was also derived from the \reff{highest} resolution \'echelle spectra \reff{among our targets}, along with a value for \Teff. The resulting parameters reported in Tables \ref{tab:star1} and \ref{tab:star2} for this object may therefore be considered the most reliable.

Reliability as well as precision in the measurement of the fundamental properties of M~dwarfs is important in order to address uncertainties surrounding the interior physics governing the evolution of them. Fig.\,\ref{fig:mass-radius} displays the objects plotted in the mass-radius plane along with all M~dwarfs with radii previously determined to better than 10$\%$ as catalogued by \citet{Parsons_2018}. The figure also displays BCAH15 isochrones for 0.05, 0.1, 0.3 Gyr and 10 Gyr. Notice the rapid contraction between 0.05 and 0.3 Gyr as the model evolves toward the main sequence. In the main sequence, between 0.3 and 10 Gyr, theory systematically underestimates M~dwarf radii. This conclusion is verified in Fig.\,\ref{fig:mass-log(g)}, where surface gravity is used instead of radius; the surface gravity of the M-dwarf is derived from the light-curve and RVs alone \citep{Southworth_2007}, so is not dependent on the models. We also show the same objects in the mass-\Teff\ plane with the same isochrones. The SB2 systems suggest that temperatures are overestimated by the models. We note that our M-dwarf \Teff\ predictions appear to conform to the general trend set by SB2 M-dwarfs better than other SB1 M-dwarf determinations for this parameter.

\begin{figure}
    \centering
    \includegraphics[width = 8.5cm]{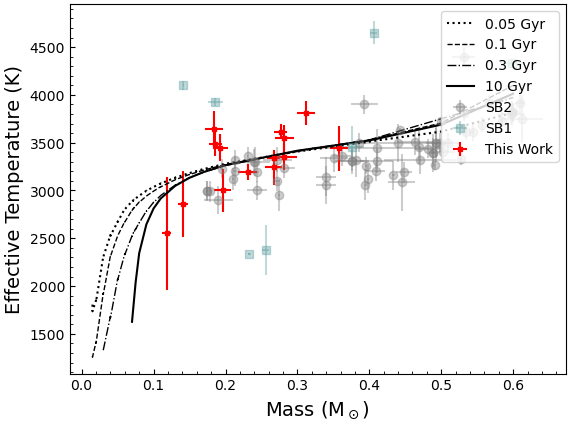}
    \caption{BCAH15 isochrones for 0.05, 0.1, 0.3 and 10 Gyr plotted in the mass-\Teff\ plane with the locations of the current objects over-plotted in red. M~dwarfs from \citet{Parsons_2018} are also shown where SB2 M-Dwarfs are plotted in grey and SB1 M-dwarfs are plotted in blue.}
    \label{fig:mass-teff}
\end{figure}

\begin{figure}
    \centering
    \includegraphics[width = 8.5cm]{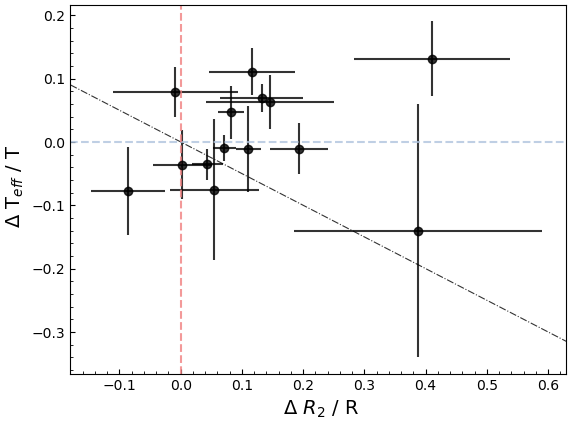}
    \caption{The \reff{fractional} radius and effective temperature discrepancies relative to theoretical predictions of BCAH15. Below the dashed blue line, objects show overestimated temperatures by the models for the corresponding age and to the right of the red dashed line, objects are inflated. \reff{The constant luminosity hypothesis is represented by the black dash-dotted line.}}
    \label{fig:Delta_Corr}
\end{figure}

It has been claimed that a correlation exists where inflated radii are accompanied by a cooler \Teff\ such that luminosity is unaffected \citep[e.g.][]{Torres07apj}. We investigated this by plotting the two discrepancies relative to BCAH15 isochrones for the objects studied in this work in Fig.\,\ref{fig:Delta_Corr}. The discrepancies were calculated relative to the corresponding isochrones for each object's age, so this was not possible for the rest of the objects in Fig.\,\ref{fig:mass-radius}. \reff{Below the dashed blue line, objects show overestimated \Teff s by the models, and to the right of the red dashed line, objects are inflated, i.e., underestimated by the models. It does not appear that inflated radii are accompanied by an overestimated \Teff\ from this sample. This is in contrast to the majority of SB2 systems that do show overestimated \Teff s by the models in Fig.\,\ref{fig:mass-teff}. It is possible that different physical processes due to the brighter companion, such as global redistribution of a larger amount of incident radiation, may affect the adequacy of using SB1 systems to test the constant luminosity hypothesis, and this could be why the radius, as well as the \Teff, is underestimated by the models for five of the objects in this study. Furthermore, the constant luminosity hypothesis requires a gradient of $-0.5$, which is represented by the black dash-dotted line in Fig.\,\ref{fig:Delta_Corr}. We found that removing the outlier, GSC-06465-00602 (bottom right in the figure), from the investigation yields a linear fit to the remaining objects that satisfied a gradient of $+0.42$; the weighted Pearson correlation coefficient of 0.217, however, indicates that this is only a weak correlation.} 

TYC 3700-1739-1B shows the largest amount of inflation compared to the models but note that this is a young system, as estimated by the previous authors as well as the current study. The margin of uncertainty in the current age estimate for this system encompasses the entire pre-main sequence phase, where the models show the largest and most rapid variation in radius. Taking the lower limit of the age estimation in the determination of its inflation status leads to the conclusion that the object is under-inflated after previously being stated to be inflated by $41\%$. It is therefore vital that precise age estimates are obtained for young stars in order to be able to accurately compare their properties to theory. The same scenario is observed for GSC 06465-00602, where it is observed to be inflated by a large amount while the uncertainty in its age spans the pre-main sequence. It is concluded that the inflation status of an M~dwarf cannot be accurately addressed if the uncertainty in its age estimate allows a pre-main-sequence solution, where small changes in the object's age would change the result. This necessity for precise age estimations due to rapid evolutionary changes makes these young objects valuable tests for our theoretical understanding.

Large uncertainties in age estimations are common but this has a much smaller effect on the predicted radius of a main sequence M~dwarf. Inflation observed well within the main sequence is therefore likely accurate. The contraction of the core throughout the main sequence due to an increasing mean molecular weight results in a slight increase in the star's radius but most M~dwarfs are found to be inflated beyond this. This is demonstrated in Fig.\,\ref{fig:mass-radius} by the gap between the 0.3 and 10 Gyr BCAH15 isochrones, where most M~dwarfs lie above it. 

A widely discussed hypothesis is related to the binarity of the majority of M~dwarfs with measured radii precise to $10\%$. Increased rotational velocity induced via synchronization between the orbital and rotational periods speeds up the internal dynamo, increasing magnetic activity and decreasing the efficiency of convection \citep{Fernandez_2009}. This may cause the radius to expand \citep{Mullan_2001, Lopez-Morales_2007}. However, the dynamos may operate differently in fully convective stars \citep{Zhou_2014} so the relationship between rotation and activity regarding such systems and its relevance to radius inflation is unclear. As stated in Section~\ref{Introduction.S}, the explanation needs not to be restricted to binary stars due field stars measured via interferometry demonstrating the same discrepancy compared to models; field stars rotate slowly due to magnetic braking.

\citet{Berger_2006} found the disagreement to be larger for metal rich stars, concluding that an opacity component missing from the models may lead to larger radii for stars with larger metallicity. Then it is interesting that \reff{two} of the M~dwarfs (TYC 6493-290-1 and GSC 06493-00315), found to agree with their isochrones in the current work, were previously found to have sub-solar metallicity by their previous authors, \citet{Zhou_2014} and \citet{Fernandez_2009}. It is of particular interest to determine whether the remaining M~dwarf which agrees with isochrones, HAT-TR-205-003, is also metal-poor. This would support the hypothesis of \citet{Berger_2006} and other authors who have suggested the treatment of metallicity in the models as a source for inflation.

On the other hand, \citet{vonBoetticher} found a correlation between radius residual with solar isochrones and estimated metallicity, such that for objects with values for [Fe/H] $\neq 0$, accounting for it in the isochrones would act to reduce the observed inflation or remove it. Estimations for [Fe/H] derived from high resolution \'echelle spectra would therefore be particularly valuable because it would mean that the radius residuals against the solar isochrones can be corrected for it as well as being able to include the value for [Fe/H] in the derivation of the empirical M dwarf radius. \reff{We note, since GSC-06465-00602 is the only inflated object in the current study with a significantly sub-solar, and potentially reliable value for metallicity, the reliability of the investigation into the size of the discrepancy might be improved by using a non-solar metallicity isochrone, rather than the solar BCAH15 isochrones} 

GSC 05946-00892 is the only object in the current study found to be under-inflated compared to models. This object also has the largest surface gravity estimation among the other objects included in this work.

We have used the BCAH15 isochrones to determine how inflated the M~dwarfs are in each system. However, these models may have a different age scale from the models used to determine the properties of the primary stars. The difference between the age scales is at most 0.2\,Gyr, so should not affect our conclusions.

\section{Conclusion}

We have presented determinations of the physical properties of a set of 15 EBLMs using TESS and new ground-based light curves plus published RVs. Photometric data were modelled simultaneously with the RVs using \textsc{jktebop} and physical parameters were calculated using an isochrone fitting method, yielding masses and radii of both components as well as the orbital semimajor axis and age of the systems. Our results are the first measurements of these properties for two of the systems. Our results improve and extend the catalogue of available physical properties of low-mass stars.

The full phase coverage of the TESS light curves means that the secondary eclipse for 14 of these objects has been observed and analysed for the first time, allowing for an estimation of the surface brightness ratio and \Teff\ of the M~dwarf, whilst also reliably constraining the eccentricity of the systems. Our M~dwarf \Teff\ predictions appear to be more reliable than previous attempts to estimate this value for M~dwarfs in SB1 systems.

Estimated \Teff\ values for the M~dwarfs allowed for the objects to be displayed in the mass-\Teff\ plane as well as the mass-radius diagram, among other well-characterised M~dwarfs, and discussed in the context of radius inflation. It was discovered that exquisite precision in the age estimate of young stars is required in order to reliably address their inflation status. \reff{Neglecting such objects (TYC 3545-371-1, TYC 3700-1739-1) from the following statistic due to the uncertainty in the determination of its inflation status, 10 out of the 13 remaining objects were found to be inflated, by 11.4$\%$ on average. We do not find evidence from our SB1 sample of M-dwarfs that luminosity is unaffected by inflation; however, we note that our sample of objects with inflated radii is relatively small compared to the amount of SB2 M dwarfs in Fig. \ref{fig:mass-teff} where \Teff\ is overpredicted by models.} 

Precise measurements of the metallicity and \Teff\ for these systems would improve the reliability of the results and \reff{possibly} remove ambiguity regarding \reff{some of the} disagreements between the current and previous results. These would ideally be based on new high-quality \'echelle spectra. Additional RV measurements would be useful for many of the systems, and TESS continues to observe the objects we have studied. Our work is therefore an important improvement, but not the final word, in our understanding of these objects.

\section*{Data availability}

This paper includes data collected by the TESS mission which is publicly available at the Mikilski Archive for Space Telescopes (MAST) at the Space Telescope Science institute (STScl) (\href{https://mast.stsci.edu/portal/Mashup/Clients/Mast/Portal.html}{https://mast.stsci.edu}).

This paper includes photometry collected using the CAHA 2.2-m using telescope time awarded under the OPTICON Transnational Access scheme via proposal 2010B/02, and the CAHA 1.23-m telescope. These data will be made available at the Centre de Donn\'ees astronomiques de Strasbourg (CDS, \href{http://cdsweb.u-strasbg.fr/}) upon publication of this paper.

This paper uses data provided by previous authors of the targets studied here and is publicly available at those references unless stated that the data were provided via private communication.

\section*{Acknowledgements}

Based on observations collected at the Centro Astron\'omico Hispano en Andalucía (CAHA) at Calar Alto, operated jointly by Junta de Andaluc{\'\i}a and Consejo Superior de Investigaciones Cient{\'\i}ficas (IAA-CSIC). The research leading to these results has received funding from the European Community's Seventh Framework Programme (FP7/2007-2013) under grant agreement number RG226604 (OPTICON).

We gratefully acknowledge financial support from the Science and Technology Facilities Council.
This research has made use of the SIMBAD and CDS databases operated by the Centre de Donn\'ees astronomiques de Strasbourg, France.
We also made use of data from the European Space Agency (ESA) mission Gaia (https://www.cosmos.esa.int/gaia), and TESS. Funding for the TESS mission is provided by NASA’s Science Mission directorate. We acknowledge the use of public TESS Alert data from pipelines at the TESS Science Office and at the TESS Science Processing Operations Center.

We are grateful to Songhu Wang and Nicolas Crouzet for providing us with their RV measurements for TYC 9535-351-1 via private communication. 
We thank Amaury Triaud for useful comments on a draft of this work.

\bibliographystyle{mnras}
\bibliography{refs.bib}

%%%%%%%%%%%%%%%%%%%%%%%%%%%%%%%%%%%%%%%%%%%%%%%%%%

%%%%%%%%%%%%%%%%% APPENDICES %%%%%%%%%%%%%%%%%%%%%

\appendix
\section{FFI DATA}
\begin{table} \caption{\label{tab:FFI_lightcurves}FFI lightcurves. }
\begin{tabular}{llll}
\hline
Object & BJD & Flux & Flux Error \\
\hline
HAT-TR-205-003  &  2458739.077576  &  0.99873  &  0.014054  \\
HAT-TR-205-003  &  2458739.09845  &  0.99998  &  0.014046  \\
HAT-TR-205-003  &  2458739.119263  &  0.998557  &  0.014051  \\
HAT-TR-205-003  &  2458739.140076  &  0.996978  &  0.014061  \\
HAT-TR-205-003  &  2458739.16095  &  0.997383  &  0.014057  \\
...\\
TYC-3121-1659-1   &  2458683.368347  &  1.002666  &  0.030907  \\
TYC-3121-1659-1   &  2458683.389221  &  1.000558  &  0.031022  \\
TYC-3121-1659-1   &  2458683.410034  &  1.000909  &  0.030932  \\
TYC-3121-1659-1   &  2458683.430847  &  0.999221  &  0.030956  \\
TYC-3121-1659-1   &  2458683.451721  &  1.000576  &  0.030973  \\
...\\
TYC-9535-351-1   &  2458658.097229  &  1.00047  &  0.005033  \\
TYC-9535-351-1   &  2458658.118042  &  1.000111  &  0.005034  \\
TYC-9535-351-1   &  2458658.138855  &  1.000515  &  0.005033  \\
TYC-9535-351-1   &  2458658.159729  &  1.000013  &  0.005033  \\
TYC-9535-351-1   &  2458658.180542  &  1.000455  &  0.005034  \\
...\\
TYC-6493-290-1    &  2458440.118758  &  1.000783  &  0.016523  \\
TYC-6493-290-1    &  2458440.139592  &  1.000544  &  0.016509  \\
TYC-6493-290-1    &  2458440.160426  &  0.999268  &  0.016512  \\
TYC-6493-290-1    &  2458440.18126  &  0.999072  &  0.016512  \\
TYC-6493-290-1    &  2458440.202094  &  0.999967  &  0.016521  \\
...\\
GSC-06493-00315   &  2458438.931227  &  0.998997  &  0.03829  \\
GSC-06493-00315   &  2458438.952061  &  0.997272  &  0.038263  \\
GSC-06493-00315   &  2458438.972895  &  0.999705  &  0.038336  \\
GSC-06493-00315   &  2458438.993729  &  1.000201  &  0.03831  \\
GSC-06493-00315   &  2458439.014562  &  0.999685  &  0.038296  \\
...\\
GSC-05946-00892   &  2458468.285434  &  0.998046  &  0.016077  \\
GSC-05946-00892   &  2458468.306268  &  1.000788  &  0.016086  \\
GSC-05946-00892   &  2458468.327102  &  0.998412  &  0.016075  \\
GSC-05946-00892   &  2458468.347936  &  0.999939  &  0.016085  \\
GSC-05946-00892   &  2458468.36877  &  1.002212  &  0.016088  \\
...\\
GSC-06465-00602  &  2459174.250599  &  1.003998  &  0.022081  \\
GSC-06465-00602  &  2459174.257543  &  1.002118  &  0.022026  \\
GSC-06465-00602  &  2459174.264488  &  0.998189  &  0.022001  \\
GSC-06465-00602  &  2459174.271432  &  0.999789  &  0.022079  \\
GSC-06465-00602  &  2459174.278376  &  1.002672  &  0.022037  \\
...\\
TYC-3700-1739-1   &  2458791.765272  &  1.000804  &  0.005028  \\
TYC-3700-1739-1   &  2458791.786106  &  1.000717  &  0.005026  \\
TYC-3700-1739-1   &  2458791.80694  &  1.000751  &  0.005031  \\
TYC-3700-1739-1   &  2458791.827773  &  1.001269  &  0.005035  \\
TYC-3700-1739-1   &  2458791.848607  &  1.000655  &  0.005032  \\
\hline
\end{tabular} \end{table}

%%%%%%%%%%%%%%%%%%%%%%%%%%%%%%%%%%%%%%%%%%%%%%%%%%

% Don't change these lines
\bsp	% typesetting comment
\label{lastpage}
\end{document}